\documentclass[aps,pre,twocolumn,superscriptaddress,showpacs,showkeys]{revtex4-1}
\usepackage{color}
\usepackage{amssymb}
\usepackage{enumerate}
\usepackage{amsmath}
\usepackage{bm}
\usepackage{graphicx}
\usepackage{newtxtext,newtxmath}
\usepackage{kotex}
\usepackage{xfrac}
\usepackage{amsfonts}

\newcommand{\kmin}{{k_o}}
\newcommand{\xmin}{{x_o}}

\newcommand{\pdegconfig}{{D}}
\newcommand{\pdegstatic}{{D_{\rm static}}}

\newcommand{\polylog}{{\rm Li}}

\begin{document}
\setcounter{page}{1}
\title{{Giant component in a configuration-model power-law graph with a variable number of links }}
\author{Heung Kyung   \surname{Kim}}
\affiliation{Department of Physics, Inha University, Incheon 22212, Korea}
\author{Mi Jin  \surname{Lee}}
\affiliation{Department of Physics, Inha University, Incheon 22212, Korea}
\author{Matthieu \surname{Barbier}}
\affiliation{Centre for Biodiversity Theory and Modelling, Theoretical and Experimental Ecology Station, CNRS, 09200 Moulis, France}
\author{Sung-Gook \surname{Choi}}
\author{Min Seok \surname{Kim}}
\author{Hyung-Ha \surname{Yoo}}
\author{Deok-Sun \surname{Lee}}
\email{deoksun.lee@inha.ac.kr}
\affiliation{Department of Physics, Inha University, Incheon 22212, Korea}

\begin{abstract}

We generalize an algorithm used widely in the configuration model such that power-law degree sequences with the degree exponent $\lambda$ and the number of links per node $K$ controllable independently may be generated. It yields the degree distribution in a different form from that of the static model or under random removal of links while sharing the same $\lambda$ and $K$. With this generalized power-law degree distribution,  the critical point $K_c$ for the appearance of the giant component remains zero not only for $\lambda\leq 3$ but also for $3<\lambda<\lambda_l \simeq 3.81$. This is contrasted with $K_c=0$ only for $\lambda\leq 3$ in the static model and under random link removal. The critical exponents and the cluster-size distribution for $\lambda<\lambda_l$ are also different from known results. By analyzing the moments and the generating function of the degree distribution and comparison with those of other models,  we show that the asymptotic behavior  and the degree exponent may not be the only properties of the degree distribution relevant to the critical phenomena but that its whole functional form can be relevant. These results can be useful in designing and assessing the structure and robustness of networked systems.
\end{abstract}

\date{\today}
\maketitle


\section{Introduction}

An important discovery made recently for the structure of complex systems is the universal broad distribution of degree, the number of the nearest neighbors~\cite{barabasi99,Faloutsos:1999:PRI:316194.316229,Albert:1999aa,Jeong:2000kx}. A number of computational models and algorithms~\cite{aiello2001,PhysRevE.64.026118,goh01,PhysRevLett.89.258702} have been proposed for implementing the power-law (PL) graphs which have the degree distribution decaying as a power law, $\pdegconfig(k)\sim k^{-\lambda}$ for large $k$, and have been instrumental in the study of the impact of this new class of disorder in percolation~\cite{PhysRevLett.85.4626,lee04}, the Ising model~\cite{PhysRevE.66.016104}, epidemic spreading~\cite{PhysRevLett.86.3200}, synchronization~\cite{PhysRevE.70.026116}, boolean dynamics~\cite{aldana03},  and many other areas~\cite{dorogovtsev08,bbv}. 

Among the remarkable results is  the zero critical point appearing when the degree exponent $\lambda$ is equal to or smaller than $\lambda_l=3$~\cite{dorogovtsev08,bbv}.  $\lambda_l$ can be called the lower-critical degree exponent in the sense that no phase transition occurs for $\lambda\leq \lambda_l$, similarly to the lower-critical dimension~\cite{goldenfeld1992lectures}. The divergence of the second moment of the  degree distribution for $\lambda\leq 3$ leads to such a zero critical point in various dynamical processes, while there are exceptions such as the susceptible-infected-susceptible model for which the zero critical point is observed for all finite $\lambda$~\cite{PhysRevLett.111.068701}. 
Due to the diverging  second or higher moments of the PL degree distributions,  the critical exponents vary continuously  with $\lambda$ when $\lambda$ is smaller than the upper-critical degree exponent $\lambda_u$, which is known to be  $4$ in percolation~\cite{PhysRevE.66.036113} and $5$ in the Ising model~\cite{PhysRevE.66.016104} or the Kuramoto model for synchronization~\cite{PhysRevE.72.026208}. For $\lambda>\lambda_u$, the critical exponents take the mean-field values.

Given such a crucial role of the degree exponent  in the critical phenomena on PL graphs, a natural question arises: Is the large-$k$ behavior  the {\it only} property of the degree distribution $D(k)$ relevant to critical phenomena? There are various relevant factors beyond the degree distribution, such as degree-degree correlation~\cite{PhysRevE.76.026116} or the spectral dimensions~\cite{PhysRevE.77.036115,PhysRevLett.109.088701,Yoo2018},  but we are here focused on whether different degree distributions sharing the same degree exponent could lead to different critical phenomena from  described  above and how much different if they do. The answers  are not immediately clear, comparing critical phenomena on the PL graphs with degree distributions in different form but sharing the same degree exponent.

In this paper, we consider the formation and growth of the giant connected component in random PL graphs of $N$ nodes and degree exponent $\lambda$ as the number of links per node $K=L/N$ increases, called the percolation problem,  and investigate how the functional form of the degree distribution affects the critical phenomena.  To this end, model PL graphs of different values of $K$ for given $\lambda$ should be constructed, which have been so far done by (1) removing links randomly in a graph of sufficiently many links~\cite{PhysRevLett.85.5468,PhysRevLett.85.4626}  or (2) adding links one by one to connect node pairs stochastically under prescribed inhomogeneous connection probabilities ~\cite{goh01}. The latter, called the static model, generalizes the Erd\H{o}s-R\'{e}nyi graphs constructed with a uniform connection probability~\cite{Erdos:1960}. In both cases, the analytic expressions for the size of the giant component and the size distribution of  finite clusters are available~\cite{PhysRevE.64.026118,PhysRevE.66.036113,lee04},  and reveal the critical point and the critical exponents as sketched above. 

Here we consider the configuration-model PL graphs~\cite{BENDER1978296,Molloy:1995aa,aiello2001,PhysRevE.64.026118,PhysRevE.71.027103} for arbitrary $K$ and $\lambda$ with the degree distribution in a different form from  (1) or (2) above.  
The degree distribution that we propose is not an arbitrary one but followed by  the PL degree sequences generated by  a modified version of the algorithm easy to implement and thus adopted in a popular computation library~\cite{configmodel_networkx} used widely in network research. 
The modification allows the PL degree sequences to have an arbitrary value of  $K$ for  given $\lambda$, which is not possible with the original algorithm but crucial for the study of the percolation problem. 

The lower-critical degree exponent $\lambda_l$ is  shown to be $3.8106\ldots$ with this generalized degree distribution.  It is larger than $3$. Moreover, for $2<\lambda<\lambda_l$,  the giant component grows linearly with $K$ for small $K$ and the cluster-size distribution decays exponentially,  which are contrasted with the super-linear growth of the giant component and the PL decay of the cluster-size distribution in the static model.  The moments and the generating function of the degree distribution are analyzed to understand the origin of these phenomena, which leads us to see that not only the large-$k$ behavior but also the whole functional form of the degree distribution $D(k)$ is relevant to the critical phenomena.    The density of low-degree nodes, which have received relatively little attention,  is shown to be crucial in the approximate expression for the giant component size in the supercritical regime.

This paper is organized as follows. In Sec.~\ref{sec:model}, the algorithm of generating the PL degree sequence with tunable number of links per node  and its degree distribution is introduced and the basic properties are analyzed.  The generating function method is applied with the proposed degree distribution to derive the critical behaviors of the  giant component's size  and the cluster-size distribution, which  are compared with other models in Sec.~\ref{sec:size}. We investigate the giant component size in the supercritical regime and show the important roles  of low-degree nodes in model and real-world networks in Sec.~\ref{sec:low}. Our findings are summarized and discussed in Sec.~\ref{sec:summary}.


\section{Generating power-law degree sequences with tunable number of links per node: A generalized degree distribution}
\label{sec:model}

\subsection{Brief introduction of the static model}
In the static model~\cite{goh01}, links are added one by one to node pairs  selected under a prescribed probability.  In the configuration model~\cite{BENDER1978296,Molloy:1995aa,aiello2001,PhysRevE.64.026118,PhysRevE.71.027103}, on the other hand, the degree of each node is first determined from a prescribed degree distribution and then the link stubs are randomly paired. Before presenting our new degree distribution for the configuration model and investigating its properties, we briefly introduce the static model first as its properties are compared with those of our configuration model throughout this paper. 

 In the static model for constructing a PL graph with $N$ nodes, $L= N K$ links, and degree exponent $\lambda$, each node $i$ is assigned the selection probability $w_i = {i^{-\alpha}\over \sum_{\ell=1}^N \ell^{-\alpha}}$  with $\alpha = {1\over \lambda-1}$.  Two nodes $i$ and $j$ are selected with probability $w_i w_j$ and are connected if they are disconnected. This procedure is repeated until $L$ distinct pairs of nodes are connected. As distinct node pairs are connected independently, various properties  are accessible analytically. For instance, the  degree distribution is obtained as~\cite{lee04}
\begin{align}
\pdegstatic(k) &= {1\over k!} {d^k \over dz^k} \tilde{\Gamma} \left(2K (1-z)\right),
\label{eq:Dkstatic}
\end{align}
where $\tilde{\Gamma}(y) = (\lambda-1) \left({\lambda-2\over \lambda-1} y \right)^{\lambda-1} \Gamma\left(1-\lambda, {\lambda-2 \over \lambda-1} y\right)$ with $\Gamma(s,x)$ the incomplete Gamma function $\Gamma(s,x)\equiv \int_x^\infty dt \, t^{s-1} e^{-t}$. It  takes a PL form $\pdegstatic(k) \sim k^{-\lambda}$ for large $k$ and some examples are shown in Fig.~\ref{fig:KxminPk}. The size of the giant component can be obtained analytically as a function of the number of links per node $K$~\cite{lee04}, which is summarized in Appendix~\ref{sec:static}.

\subsection{Configuration model with a generalized degree distribution}
\label{sec:model_config}
For constructing PL graphs in the configuration model~\cite{BENDER1978296,Molloy:1995aa,aiello2001,PhysRevE.64.026118,PhysRevE.71.027103}, a degree sequence $\{k_1, k_2, \ldots, k_N\}$ is first generated and assigned to nodes such that each node $i$ has $k_i$ link stubs. Then randomly selected pairs of stubs from distinct nodes are connected, avoiding multiple links, which is repeated until no isolated stub is left. The degree sequence is generated by drawing a random number $k$ from a desired degree distribution $D(k)$. Therefore one can construct a PL graph by generating a degree sequence from 
\begin{equation}
\pdegconfig(k) = {k^{-\lambda} \over \zeta(\lambda)} \ \  (k\geq 1),
\label{eq:dk0}
\end{equation}
where $\zeta(\lambda)$ is the Riemann zeta function. 

Random integers $\{k_i |i=1,2,\ldots, N\}$ following $D(k)$ in Eq.~(\ref{eq:dk0}) can be generated in various ways including  the rejection method~\cite{PhysRevE.64.026118}, the Walker algorithm~\cite{lee04}, and the transformation method, e.g., rounding real random numbers $X$ following a Pareto distribution 
\begin{equation}
F (x) = {\rm Prob.} (X\geq x) = 
\begin{cases}
1& (x<1)\\
x^{1-\lambda}  &  (x\geq 1)
\end{cases}
.
\label{eq:pareto}
\end{equation}
The transformation method is easy to implement, as  a random number $r$ uniformly distributed between $0$ and $1$ can give $X$ via the relation
$X = (1- r)^{-{1\over \lambda-1}}$,  and is therefore adopted  in the widely used python libraries {\it random} and {\it NetworkX}~\cite{configmodel_networkx}. 

Given the ubiquity of PL degree distributions in complex systems, it is an advantage of the configuration model that graphs with such a perfect power law as in Eq.~(\ref{eq:dk0}) can be generated. Indeed, the model has been widely used in the study of the structure and dynamics of complex networks.  Yet there is a flaw in Eq.~(\ref{eq:dk0}).  The number of links per node $K=L/N$ cannot be varied freely but fixed for given degree exponent $\lambda$: $K = \langle k\rangle/2 = \sum_k k \, D(k) = {\zeta(\lambda-1)\over 2\zeta(\lambda)}$. The expected size of the giant component is  also  fixed for given $\lambda$, which prevents us from studying the evolution of the giant component with increasing $K$ for given $\lambda$. Instead, the percolation problem has been studied as $\lambda$ varies, yielding  $\lambda_c \simeq 3.47875$~\cite{aiello2001}. One may assume that Eq.~(\ref{eq:dk0}) applies for $k\geq k_{\rm min}$ with $k_{\rm min}$ an integer, but even then $K$ cannot take arbitrary values but takes just selected discrete values: $K = {\zeta(\lambda-1,k_{\rm min})\over 2\zeta(\lambda,k_{\rm min})}$ for different $k_{\rm min}$ with $\zeta(s,a)$ the Hurwitz zeta function. 
To overcome this limitation, we generate a PL degree sequence using the following algorithm.

\medskip

\noindent (1) For a node $i$, draw a random variable $r$ between $0$ and $1$ from the uniform distribution and obtain $X = \xmin \, (1-r)^{-{1\over \lambda-1}}$ with $\xmin$ a positive real constant. $X$ is a real-valued random number in the range $X>\xmin$ and follows the Pareto distribution 
\begin{equation}
F_{\xmin} (x) = {\rm Prob.} (X\geq x) =
\begin{cases}
1 & (x<\xmin)\\
 \left({x\over \xmin}\right) ^{1-\lambda}  & (x\geq \xmin)
 \end{cases}
 .
\label{eq:pareto_xmin}
\end{equation}
(2) Take the integer part of $X$ to get the degree of node $i$ as 
\begin{equation}
k_i = \lfloor X\rfloor
\label{eq:kix}
\end{equation}
with $\lfloor x\rfloor$ the largest integer not larger than $x$. \\
(3) Repeat (1) and (2) for every node $i=1,2,\ldots, N$ to obtain a PL degree sequence $\{k_1, k_2, \ldots, k_N\}$. 

\medskip

Here $\xmin$ is a parameter allowing us to control the  number of links per node $K$, and the lowest degree is given by $\kmin = \lfloor \xmin\rfloor$.  In simulations, we further restrict the maximum degree, but we do not discuss this restriction in the following, as its effects~\cite{PhysRevE.71.027103} are not relevant to the results presented in this paper (see Appendix~\ref{sec:uncorr}). 

 The degree distribution $\pdegconfig_{\xmin,\lambda}(k)$  is then equal to the probability that $X$ is between $k$ and $k+1$, evaluated as 
\begin{align}
\pdegconfig_{\xmin} (k)
&= F_{\xmin} (k) - F_{\xmin}(k+1) \nonumber\\
&= 
\begin{cases}
1 - \xmin^{\lambda-1} (\kmin +1)^{1-\lambda} & {\rm for} \ k = \kmin,\\
\xmin^{\lambda-1} \left[ k^{1-\lambda} - (k+1)^{1-\lambda}\right] & {\rm for }\   k>\kmin
\end{cases}
.
\label{eq:Dkconfig}
\end{align}
Notice that the degree distribution behaves as a power law
\begin{equation}
\pdegconfig_{\xmin} (k) \simeq (\lambda-1) \xmin^{\lambda-1} k^{-\lambda} \ \  {\rm for} \  k\gg 1.
\end{equation}
\begin{figure}
\includegraphics[width=\columnwidth]{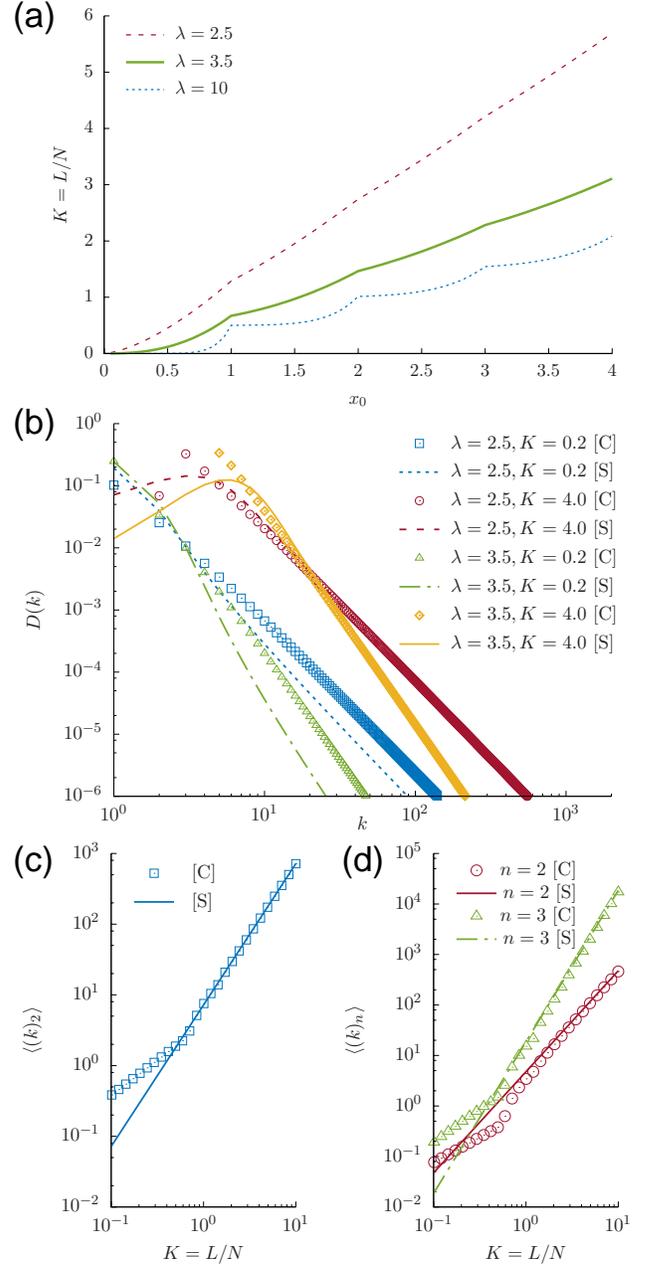}
\caption{The degree distribution and its moments for the PL degree sequence  from the proposed algorithm. (a) The number of links per node $K=L/N$ as a function of the parameter $\xmin$ as given in Eq.~(\ref{eq:Kxmin})  for different degree exponents $\lambda$. (b) The degree distribution $D(k)$ in Eq.~(\ref{eq:Dkconfig}) of the configuration-model graphs ([C])  based on our PL degree sequence for different $\lambda$ and $K$ (points). For comparison, the degree distribution of the static model ([S]) from Eq.~(\ref{eq:Dkstatic}) is shown (lines). 
(c) The second factorial moments $\langle (k)_2\rangle = \langle k(k-1)\rangle$ as a function of $K=L/N$ in our configuration-model graphs ([C]) and  the static-model graphs ([S]) for $\lambda=3.5$. (d) Plots of the second and third factorial moments $\langle (k)_2\rangle$ and $\langle (k)_3\rangle=\langle k(k-1)(k-2)\rangle$  versus $K$ for $\lambda=4.5$. }
\label{fig:KxminPk}
\end{figure}
The number of links per node $K$ can be changed by varying the parameter $\xmin$ as
\begin{align}
2 K &= \langle k\rangle  = \sum_k k\, \pdegconfig_\xmin (k) 
=\kmin + \xmin^{\lambda-1} \zeta_\infty(\lambda-1,\kmin+1),
\label{eq:Kxmin}
\end{align}
where     $\zeta_\infty (\lambda,a)  \equiv \sum_{k=a}^\infty k^{-\lambda}$ is identical to the Hurwitz zeta function $\zeta(\lambda,a)$ for $\lambda\geq 1$;  Note that $\zeta_\infty(s,a)$ diverges for $s<1$ but $\zeta(s,a)$ is finite for $s<1$  by analytic continuation. We will restrict ourselves to the range $\lambda>2$ to avoid the case of diverging $K$. Also we denote $\zeta_\infty (s,1)$ by $\zeta(s)$ if $s\geq 1$. 

As shown in Fig.~\ref{fig:KxminPk} (a), $K$ increases monotonically from $0$ to infinity   with $\xmin$, which ensures the unique value of $\xmin(K)$ for given $K$. Therefore, for arbitrary $K$ and $\lambda>2$, one can construct  the configuration-model PL graphs.
Introducing the parameter $\xmin$ as in Eq.~(\ref{eq:pareto_xmin}),  the conventional algorithm  given in Eqs.~(\ref{eq:dk0}) and (\ref{eq:pareto}) has been changed  to enable us to tune  $K$ as in Eq.~(\ref{eq:Kxmin}). Moreover, as
in Fig.~\ref{fig:KxminPk} (b),   the generalized degree distribution $\pdegconfig_\xmin(k)$ has different functional form from  the static model for the same values of $K$ and $\lambda$.  The difference is more significant for smaller $K$.

In the static model and in the graphs obtained by removing links randomly, the  factorial moments  $\langle (k)_r \rangle \equiv \langle k(k-1)(k-2) \cdots (k-r+1)\rangle$ of the degree distribution  scale with respect to $K$ as $\langle (k)_r\rangle \sim K^r$ as shown in  Appendix~\ref{sec:factorial}. The factorial moments of the degree distribution in Eq.~(\ref{eq:Dkconfig}) behave differently for small $K$ as shown in Figs.~\ref{fig:KxminPk}(c) and~\ref{fig:KxminPk}(d).
The decrease of $K$ in our configuration model is realized when links are removed in such a nonrandom way that preserves the form of the degree distribution in Eq.~(\ref{eq:Dkconfig}), and therefore 
reproduces the specific scaling of the factorial moments, which is presented in 
Appendix~\ref{sec:linkremoval}. Such different behaviors of the moments between the static model and our configuration model result in  different lower-critical degree exponents for the percolation transition, which is addressed in  the next section.


\section{Critical phenomena associated with the formation of the giant component}
\label{sec:size}

When the largest connected component (LCC) of size $S$ is so large that the relative size
\begin{equation}
m=\lim_{N\to\infty} {S\over N}
\end{equation}
 is nonzero, it is called the giant component.  We investigate  the behavior of $m$ as a function of the number of links per node $K$ for a given number of nodes $N$ and degree exponent $\lambda$  in the configuration-model graphs introduced in Sec.~\ref{sec:model}. 
  This study was not possible in previous works on the configuration model  which used the degree distribution  in Eq.~(\ref{eq:dk0}) as the number of links per node is then fixed by $\lambda$.

\begin{figure}
\includegraphics[width=\columnwidth]{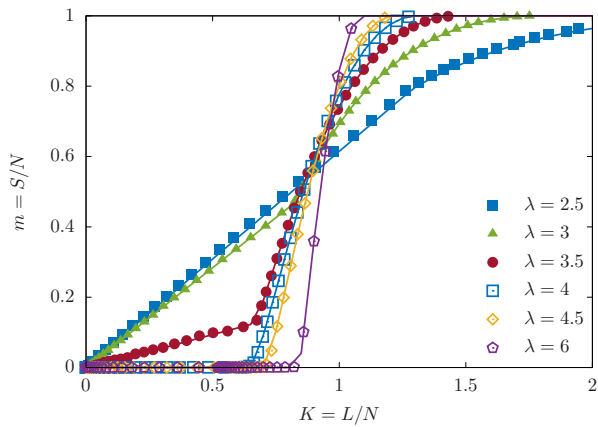}
\caption{The relative size of the LCC $m=S/N$  as a function of the number of links per node $K=L/N$ in the configuration-model PL graphs with the degree distribution in Eq.~(\ref{eq:Dkconfig}) for $N=10^6$ and selected values of the degree exponent $\lambda$. The lines are from the exact solution to Eq.~(\ref{eq:mu}), which agree very well with simulation results (points).}
\label{fig:SN}
\end{figure}

\subsection{Simulation results for the giant component's size}

The relative size of the giant component in our configuration-model graphs is given for various degree exponents in Fig.~\ref{fig:SN}. The most remarkable feature is that  up to $\lambda$ as large as $3.5$, the critical point $K_c$ at which the giant component begins to form is zero---that is, the giant component is formed for any nonzero value of $K$. 
By contrast, the static model displays a  vanishing threshold only for $\lambda\leq 3$  [see Eq.~(\ref{eq:Kc_static}) in Appendix~\ref{sec:static}]. Furthermore, when $\lambda\lesssim 3.5$, the giant component grows {\it linearly} with $K$ for small $K$ and then  abruptly changes to a concave increasing function at a certain value of $K$. In contrast, for $\lambda=4.5$ or $6$,   $m$ shows a transition behavior at  a threshold $K_c$, as in the static model. Yet the critical point $K_c$ is between $\sim 0.6$ and $1$ in our configuration-model graphs, while it is between $0$ and $1/2$ in the static model.  

These simulation results cast many questions regarding the behavior of $m$ as a function of $K$ and $\lambda$ in the configuration-model PL graphs. What is the origin of the linear growth of the giant component's size with $K$ and in what range of  $\lambda$ is that behavior observed? What is the critical point $K_c$ for large $\lambda$? How does $m$ behave near the critical point? Most of all, one may wonder whether these critical behaviors are different from the known results for the static-model PL  graphs or the random-link-removed PL graphs. It does not seem that these questions can be answered merely by examining simulation results. 

With the degree distribution in Eq.~(\ref{eq:Dkconfig}), the size of the giant component and the cluster-size distribution can be analytically obtained,  which can answer these questions. We will show that the anomalous behaviors of the giant component in our configuration model originate in specific properties of the degree distribution in Eq.~(\ref{eq:Dkconfig}), the functional form and moments of which deviate from those of the static model (Fig.~\ref{fig:KxminPk}). The  size of the giant component for small $K$ or $K$ near the critical point $K_c$ can be obtained by assuming that the giant component is of tree structure, which allows a mapping to branching processes~\cite{PhysRevLett.85.5468,lee04}. The obtained analytic solution will allow us to understand better the behavior of the giant component in our configuration-model graphs.

\subsection{Mapping to branching processes}

While the branching process approach~\cite{otter1949} for the study of cluster formation in graphs is well known and has been widely used~\cite{PhysRevLett.85.5468,PhysRevE.66.036113,lee04}, we review the method here to provide a self-contained analysis.

It can be assumed and self-consistently verified that finite clusters have a low density of loops and are almost treelike in structure~\cite{lee04}. For given $K$ and $\lambda$,  the ensemble of connected components in  realizations of these graphs can therefore be approximated by  the ensemble of trees generated by a branching process, whose branching probability is given by the degree distribution $D(k)$ of the graphs. The probability that a root node generates $k$ daughters is set to be equal to $\pdegconfig(k)$ and  the probability that a node other than the root generates $k$ daughters is given by $(k+1) \pdegconfig(k+1)/\langle k\rangle$. This mapping holds when finite connected components have a tree structure and the degrees of neighboring nodes are not correlated. Then, the cluster-size distribution $P(s)$, the probability that a node belongs to a size-$s$ cluster, corresponds to the probability of a node to be the root of a size-$s$ tree. $P(s)$ depends on the probability $R(s)$ that a link leads to a size-$s$ tree. 

Let us define the generating functions $\mathcal{P}(z) \equiv  \sum_{s<\infty} P(s) z^s$ and $\mathcal{R}(z) \equiv \sum_{s<\infty} R(s)z^s$, where the summation runs only over finite size $s$. The two generating functions satisfy the following relations
\begin{align}
\mathcal{P}(z) &= z \, g (\mathcal{R}(z)),
\label{eq:P} \\
 \mathcal{R}(z) &= z h (\mathcal{R}(z)),
\label{eq:R}
\end{align}
where  $g(z)\equiv \sum_{k=0}^\infty \pdegconfig(k) z^k$ and $h(z)\equiv \sum_{k=0}^\infty (k+1) \pdegconfig(k+1) z^{k} /\langle k\rangle = {g'(z) \over \langle k\rangle}$ are defined in terms of the degree distribution~\cite{lee04}.  Considering Eqs.~(\ref{eq:P}) and (\ref{eq:R}) at $z\to 1^{-}$ and denoting $\mathcal{R}(1)$ by $u$, we find that the giant component size is evaluated as 
\begin{align}
m &= 1 - g (u), \nonumber\\
u &= h (u).
\label{eq:mu}
\end{align}
The variable $u\equiv \mathcal{R}(1) = \sum_{s<\infty} R(s)$ represents the probability that a  link leads to a finite cluster.  $u$ is obtained by solving the second line in Eq.~(\ref{eq:mu}). The function $h(u)$ increases monotonically with $u$ from $h(0)=\pdegconfig(1)/\langle k\rangle$ to $h(1)=1$. Therefore there is always a trivial solution $u=1$. There exists a nontrivial solution $u<1$ if 
\begin{equation}
h'(1) = {\langle k^2\rangle - \langle k\rangle \over \langle k\rangle}>1,
\label{eq:g11moments}
\end{equation}
in which case the  nontrivial solution $u$ is the true value of $\mathcal{R}(1)$, and gives a nonzero value of $m$ by Eq.~(\ref{eq:mu}). Therefore the critical point $K_c$ is determined by Eq.~(\ref{eq:g11moments}), yielding the condition that the moment ratio $\langle k^2\rangle/\langle k\rangle$ must be larger than $2$ for the emergence of the giant component. Note that $h'(1)$ is equal to the ratio of the second to the first factorial moment $\langle (k)_2 \rangle/\langle k\rangle$.

\subsection{Critical point}

Let us first use Eq.~(\ref{eq:g11moments})  to determine the critical point for the giant component formation in configuration-model PL graphs with the degree distribution $\pdegconfig_\xmin(k)$  in Eq.~(\ref{eq:Dkconfig}). For given $\xmin$ and $\lambda$, the generating function $g(z)$ is given by 
\begin{align}
g(z) &= z^{\kmin} [1 - \xmin^{\lambda-1} (1-z) \Phi(z,\lambda-1,\kmin+1)], 
\label{eq:g0z}
\end{align}
and $h(z)$ is given by 
\begin{align}
h(z) &= z^{\kmin-1} {\kmin + \xmin^{\lambda-1} \Psi(z,\lambda-1,\kmin+1) \over \kmin + \xmin^{\lambda-1} \zeta_\infty(\lambda-1,\kmin+1)}, \nonumber\\
\Psi(z,s,a)&\equiv \Phi(z,s,a) - (1-z) \Phi(z,s-1,a),
\label{eq:g1z}
\end{align}
where we used Eq.~(\ref{eq:Kxmin}) for $\langle k\rangle$, $\kmin = \lfloor \xmin\rfloor$, and  the Lerch transcendent $\Phi(z,s,a) = \sum_{\ell=0}^\infty (\ell + a)^{-s} z^\ell$. 
Note that $\Phi(1,s,a) = \zeta_\infty(s,a)$ and $\Phi(z,s,1) = z^{-1} {\rm Li}_s (z)$ with ${\rm Li}_s(z) = \sum_{\ell=1}^\infty \ell^{-s} z^\ell$ the polylogarithm function.

\begin{figure}
\includegraphics[width=\columnwidth]{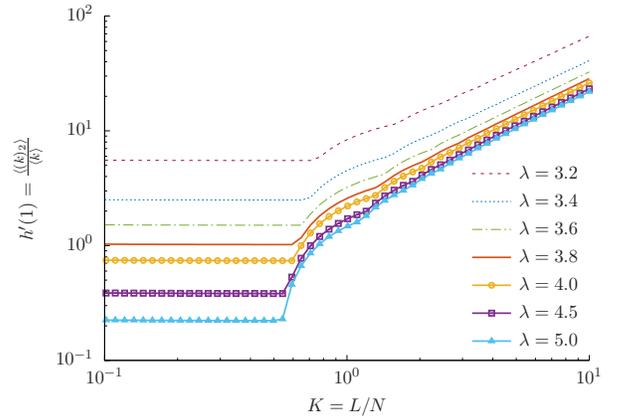}
\caption{Plot of $h'(1)$ in Eq.~(\ref{eq:g11}) as a function of the  number of links per node $K$ for different $\lambda$.} 
\label{fig:g11}
\end{figure}

Using Eq.~(\ref{eq:g1z}), we obtain 
\begin{widetext}
\begin{equation}
h'(1)= {k_o^2 - k_o + 2 \xmin^{\lambda-1} \left\{\zeta_\infty(\lambda-2,\kmin+1) -\zeta_\infty(\lambda-1,\kmin+1)\right\} \over \kmin + \xmin^{\lambda-1} \zeta_\infty(\lambda-1,\kmin+1)},
\end{equation}
\label{eq:g11}
\end{widetext}
where the relation $(\partial /\partial z) \Phi(z,s,a) = z^{-1}\{ \Phi(z,s-1,a) - a \Phi(z,s,a)\}$ is used. From Eqs.~ (\ref{eq:Kxmin}) and (\ref{eq:g11}), one can obtain $h'(1)$ as a function of $K=L/N$, which is shown in Fig.~\ref{fig:g11}. A remarkable feature is the plateau in a small-$K$ region for each $\lambda$.  More importantly, $h'(1)$ remains larger than $1$ for all nonzero $K$ as long as $\lambda\lesssim 3.8$, suggesting that the giant component forms for any nonzero $K$. 

\begin{figure}
\includegraphics[width=\columnwidth]{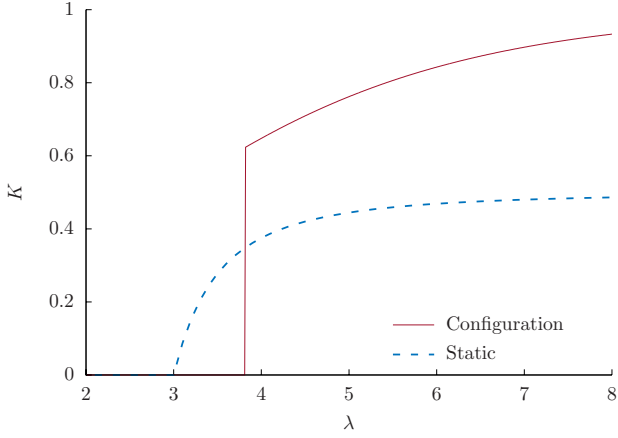}
\caption{Phase boundary between the percolating and nonpercolating phase  in the configuration model and the static model.  The critical number of links per node $K_c$  is given as a function of the degree exponent $\lambda$ such that the giant component exists for $K> K_c$.  Note that $K_c=0$ for $\lambda<3$ in the static model and for $\lambda<\lambda_l\simeq 3.81$ in the configuration model.}
\label{fig:PD}
\end{figure}

To derive analytically the condition for $h'(1)>1$, we introduce $Q(\lambda,\xmin)$ and $B(\lambda,n)$ for integer  $n$  defined as 
\begin{align}
Q(\lambda,\xmin) &\equiv  \kmin^2 - 2 \kmin + \xmin^{\lambda-1} B(\lambda,\kmin), \nonumber\\
B(\lambda,n) &\equiv 2 \zeta_\infty(\lambda-2,n+1)  - 3 \zeta_\infty(\lambda-1,n+1),
\label{eq:QB}
\end{align}
which help us see better how $h'(1)$ in Eq.~(\ref{eq:g11}) depends on $\lambda$ and $\xmin$. As $Q = \langle k\rangle \{ h'(1)-1\} = \langle k^2\rangle - 2 \langle k\rangle$, the giant component appears if $Q>0$, and the critical point is $\xmin_c$ such that $Q>0$ for $\xmin>\xmin_c$.  We now analyze the conditions of  giant component formation in the plane of $\lambda$ and $x_o$, eventually leading to the  phase diagram in Fig.~\ref{fig:PD}. 

For $2<\lambda\leq3$, the function $B(\lambda,n)$ diverges, and $Q$ is positive for any $\xmin>0$, giving the vanishing critical point $\xmin_c=0$.  In the region $\lambda>3$, $B(\lambda,n)$ decreases with increasing $\lambda$, asymptotically approaching $-1$ for $n=0$ and $0$ for $n\geq 1$ (see Appendix~\ref{sec:QB}). Therefore $Q$ can be negative only when $\xmin<2$,
and the critical point $\xmin_c$ for $\lambda>3$ should be between $0$ and $2$, if it exists. 

To further understand the behavior of $Q$ in the case of $\lambda>3$, let us first look into the region  $0<\xmin<1 (\kmin=0)$.  In this region, $Q= \xmin^{\lambda-1} B(\lambda,0)$ is positive if $B(\lambda,0)>0$, which holds for $\lambda<\lambda_l$ where $\lambda_l$ is the value of $\lambda$ satisfying the relation 
\begin{equation}
B(\lambda_l,0) = 2\zeta(\lambda_l-2)  - 3 \zeta(\lambda_l-1) = 0,
\end{equation}
and is found to be 
\begin{equation}
\lambda_l = 3.810639333567\ldots.
\label{eq:lambdac}
\end{equation}
This means that,  if $0<\xmin<1$, $Q$ is positive ($\langle k^2\rangle/\langle k\rangle>2$) in the region $3<\lambda< \lambda_l$ and negative where  $\lambda> \lambda_l$. $\lambda_l$ will be shown to be the lower-critical degree exponent below.

Next we examine the region $1\leq \xmin <2$ and $\lambda>3$. One can see that $Q = -1 + \xmin^{\lambda-1} B(\lambda,1)$ is positive as long as  $\xmin > \xmin_c(\lambda)$ with 
\begin{align}
\xmin_c(\lambda) &= B(\lambda,1)^{-{1\over \lambda-1}} \nonumber\\
&= \left\{2 \zeta(\lambda-2) - 3 \zeta(\lambda-1)+1\right\}^{-{1\over \lambda-1}}
\label{eq:xminc}
\end{align}
for $\lambda\geq \lambda_l$. 
Notice that  $\xmin_c =1$ at $\lambda=\lambda_l$ and approaches $2$ as $\lambda$ goes to infinity (Fig.~\ref{fig:xminc}).

Using $\xmin_c=0$ for $2<\lambda<\lambda_l$  and Eq.~(\ref{eq:xminc}) for $\lambda\geq \lambda_l$ and the relation between $\xmin$ and $K$ in Eq.~(\ref{eq:Kxmin}), we find that the giant component emerges  for $K>K_c(\lambda)$ with 
\begin{equation}
K_c = 
\begin{cases}
0 & {\rm for } \ 2<\lambda<\lambda_l\\
{1\over 2}\left[1 + \xmin_c^{\lambda-1} \zeta(\lambda-1, 2) \right] &\\
= {\zeta(\lambda-2)-\zeta(\lambda-1) \over 2 \zeta(\lambda-2) - 3\zeta(\lambda-1) +1} 
 & \ {\rm for} \ \lambda\geq \lambda_l
\end{cases}
,
\label{eq:Kc_config}
\end{equation}
where $\lambda_l$ and $\xmin_c$ are given in Eqs.~(\ref{eq:lambdac}) and ~(\ref{eq:xminc}), respectively.  As there is no phase transition for finite $K$ when $\lambda$ is smaller than $\lambda_l$, we call $\lambda_l$ in Eq.~(\ref{eq:lambdac}) the lower-critical degree exponent. The critical point is $K_c(\lambda_l) = (1/2)[1 + \zeta(\lambda_l-1,2)] \simeq 0.62217$ at $\lambda=\lambda_l$ and approaches $1$  for $\lambda\to\infty$, which is shown along with the critical point Eq.~(\ref{eq:Kc_static}) of the static model in Fig.~\ref{fig:PD}. The critical points for selected values of $\lambda$ in the simulation results in Fig.~\ref{fig:SN} are consistent with Eq.~(\ref{eq:Kc_config}). 

The most remarkable difference from  the known results~\cite{lee04,PhysRevLett.85.4626} is that there is a range of $\lambda$ larger than $3$ for which no phase transition occurs, although the second moment of the degree distribution does not diverge. Its origin lies in the different scaling behaviors of the moments of the degree distribution, especially the behavior of $\langle (k)_2\rangle$ with respect to $\langle k\rangle$ between our  model and the static model [Figs.~\ref{fig:KxminPk}(c) and~\ref{fig:KxminPk}(d)]. In the latter, $\langle (k)_2\rangle$ is proportional to $\langle k\rangle^2$ as shown in Eq.~(\ref{eq:kr_static}), leading $h'(1) = \langle (k)_2\rangle/\langle k\rangle$ to grow from zero linearly with $K=\langle k\rangle/2$, and consequently, $h'(1)$ exceeds $1$ at a nonzero value of $K$ as long as $\langle (k)_2\rangle$ is finite. The same is true for graphs with links removed randomly [See Eq.~(\ref{eq:kr_link})]. On the contrary, in our configuration-model PL graphs, $\langle (k)_2 \rangle$ is proportional to $\langle k\rangle$, and thus $h'(1)$ becomes a constant independent of $K$ for small $K$ as shown in Figs.~\ref{fig:KxminPk}(c) and~\ref{fig:KxminPk}(d) and Fig.~\ref{fig:g11}. 

\subsection{Critical exponent $\beta$}

The relative size $m$ of the giant component near the critical point  can be obtained analytically by solving Eq.~(\ref{eq:mu}). Let us consider the giant component size for $K$ near $K_c=0$ and $2<\lambda<\lambda_l$. When $0<\xmin<1$ or $\kmin=0$, corresponding to the number of links per node being in the range
$0<K<K_1(\lambda)$ with
\begin{equation}
K_1(\lambda) \equiv {1\over 2} \zeta(\lambda-1),
\label{eq:K1}
\end{equation}
 the function  $h(z)$ is given by 
\begin{equation}
h  (z) = {z^{-2} \left[\polylog_{\lambda-1}(z) - (1-z) \polylog_{\lambda-2}(z)\right] \over   \zeta(\lambda-1)},
\label{eq:g1z1}
\end{equation}
{\it independent} of $\xmin$ or of $K$, which gives rise to the plateaus in Fig.~\ref{fig:g11} and turns out to be responsible for the linear growth of $m$ for small $K$ and $2<\lambda<\lambda_l$ as shown below. 
Let $u_1(\lambda)$ denote the solution to $u=h(u)$ with Eq.~(\ref{eq:g1z1}) used.  
The solution $u$ to Eq.~(\ref{eq:mu})  remains fixed at $u_1(\lambda)$ as $K$ increases up to $K_1$. Then the relative size $m$ of the giant component is found to be proportional to $K$ as 
\begin{equation}
m  = 1 - g (u_1) = 2 a_{\rm (I)} K
\label{eq:m3}
\end{equation}
with the coefficient 
$
a_{\rm (I)}  = (1-u_1) {{\rm Li}_{\lambda-1} (u_1) \over u_1 \zeta(\lambda-1) }.
$
Therefore the size of the giant component grows {\it linearly} with $K$ as long as $K<K_1(\lambda)$ in Eq.~(\ref{eq:K1}) followed by a concave function $m(K)$ for $K>K_1$ (Fig.~\ref{fig:SN}).  If we define the critical exponent $\beta$ in the relation $m\sim K^\beta$ for small $K$ in case of $K_c=0$,  we find $\beta=1$ for $2<\lambda<\lambda_l$. This linear growth  is contrasted with the superlinear growth  of the giant component characterized by the exponent $\beta = 1/(3-\lambda)$ in Eq.~(\ref{eq:beta3_static}) of the static model for $2<\lambda<3$. 

The critical point $K_c$ increases from $K_c(\lambda_l)\simeq 0.62217$ towards $1$ as $\lambda$  increases from $\lambda_l$ to $\infty$, corresponding to $1<\xmin_c<2$. Let us assume that $K$ is larger than $K_c(\lambda)$ but staying around it such that $K_c(\lambda) < K<1$.  The generating function $h(z)$ with $\kmin=1$ depends on  $\xmin$ or on $K$, in contrast to the case of $2< \lambda< \lambda_l$,  and is given by
\begin{equation}
h(z) = {1 + \xmin^{\lambda-1} \left\{z^{-2} \polylog_{\lambda-1}(z) + z^{-1}(1-z^{-1}) \polylog_{\lambda-2}(z) -1 \right\} \over 1+ \xmin^{\lambda-1} \zeta(\lambda-1,2)}.
\label{eq:g1z2}
\end{equation}
Recall that its derivative at $z=1$ is larger than $1$ only for $\xmin>\xmin_c(\lambda)$ in Eq.~(\ref{eq:xminc}) or equivalently $K>K_c$ in Eq.~(\ref{eq:Kc_config}). Let us expand Eq.~(\ref{eq:g1z2})  in terms of $\alpha= -\ln z$, small around $z=1$, as 
\begin{align}
h(z &= e^{-\alpha}) \nonumber\\
&= 1 + c_1\alpha  + c_2 {\alpha^2\over 2} + \cdots + c_{\lambda-2} {\alpha^{\lambda-2} \over (\lambda-2)!} \left(1 + O(\alpha) \right),
\label{eq:g1zexpand}
\end{align}
where the coefficients are 
\begin{align}
&c_1 = -1 - {K-K_c \over K_c \left(2K_c  - 1\right)}, \nonumber\\
&c_2 = 1 +{3 \over 2} { \zeta(\lambda-3) - 3 \zeta(\lambda-2) + 2 \zeta(\lambda-1) \over \zeta(\lambda-2)  - \zeta(\lambda-1)}, \nonumber\\
&c_{\lambda-2} = {\Gamma(\lambda) \Gamma(2-\lambda) \over 2 \{ \zeta(\lambda-2) - \zeta(\lambda-1)\}},\nonumber
\end{align}
with $\Gamma(s)$ the gamma function. Using Eq.~(\ref{eq:g1zexpand}) in Eq.~(\ref{eq:mu}), we obtain the solution $u = e^{-\alpha}$ with $\alpha$ given by  
\begin{align}
\alpha \simeq 
\begin{cases}
\left\{- {c_1+1 \over c_{\lambda-2}} (\lambda-2)!\right\}^{1\over \lambda-3} = a_{\rm (II)} \Delta^{1\over \lambda-3}& {\rm for} \ \lambda_l \leq \lambda<4 \\
-2 {c_1+1 \over c_2 -1}= a_{\rm (III)} \Delta  & {\rm for } \ \lambda>4
\end{cases}
,
\label{eq:alpha}
\end{align}
with $\Delta \equiv (K/K_c -1)$, and  the coefficients 
$a_{\rm (II)} = \left[{2 \left\{ \zeta(\lambda-2) - \zeta(\lambda-1) \right\} \over \left(2K_c - 1\right)(\lambda-1) \Gamma(2-\lambda) }\right]^{1\over \lambda-3}$ and 
$a_{\rm (III)} = {4 \over 3} {\zeta(\lambda-2) - \zeta(\lambda-1) \over \left(2K_c - 1\right) \left\{\zeta(\lambda-3) - 3 \zeta(\lambda-2) + 2 \zeta(\lambda-1)  \right\} }$.
Inserting Eq.~(\ref{eq:alpha}) into $m=1 - g (u=e^{-\alpha}) \simeq 2 K_c \alpha$ with $\kmin=1$, we obtain the relative size $m$ of the giant component around the critical point $K_c$ for $\lambda\geq \lambda_l$. Using these results and Eq.~(\ref{eq:m3}), we find  that $m$ near the critical point $K_c$ follows 
\begin{equation}
m \simeq 
\begin{cases}
2 \, a_{\rm (I)}\, K & {\rm for} \ 2<\lambda<\lambda_l\\
2 \, K_c \, a_{\rm (II)} \, \Delta^{1\over \lambda-3} & {\rm for} \ \lambda_l \leq \lambda < 4\\
2 \, K_c\, a_{\rm (III)} \, \Delta & {\rm for } \lambda>4
\end{cases}
.
\end{equation}
Hence, the critical exponent $\beta$ is  
\begin{equation}
\beta = \begin{cases}
1 & {\rm for} \ 2<\lambda<\lambda_l\\
{1\over \lambda-3} & {\rm for} \ \lambda_l \leq \lambda < 4\\
1 & {\rm for} \ \lambda>4
\end{cases}
.
\end{equation}
This is different from the result obtained for the static-model PL  graphs, Eqs.~(\ref{eq:beta12_static}) and (\ref{eq:beta3_static}).  The most striking deviation is the linear growth of $m$ with $K$ even for $\lambda$ larger than $3$, up to $\lambda_l$, whereas in the static model, the giant component appears only for $K>K_c>0$ in this range of degree exponents. 
 
\subsection{Cluster-size distribution at or near the critical point}
\label{sec:cluster}

\begin{figure}
\includegraphics[width=\columnwidth]{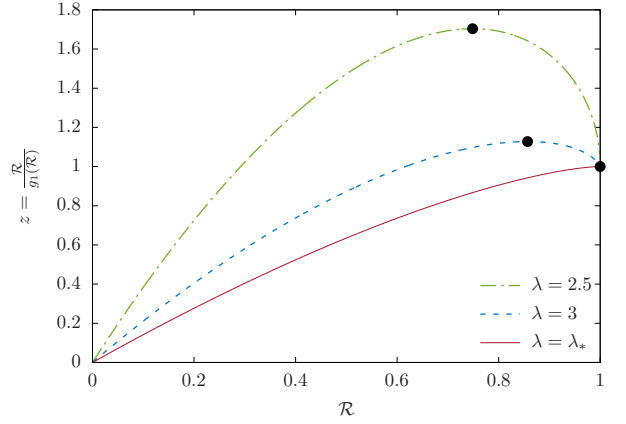}
\caption{Plots of the inverse of $\mathcal{R}(z)$, $z(\mathcal{R})=\mathcal{R}/h(\mathcal{R})$ with $\kmin=0$ for selected values of the degree exponent $\lambda$. The dots are at $(\mathcal{R}_0, z_0)$, where the derivative $dz/d\mathcal{R}$ is zero. For example, $(\mathcal{R}_0,z_0)=(0.749,1.70)$ and $(0.857,1.13)$ for $\lambda=2.5$ and $3.0$, respectively. It converges to $(1,1)$ as $\lambda$ approaches $\lambda_l$. Note that only the branch in the region $\mathcal{R}<\mathcal{R}_0$ and $0\leq z<z_0$ corresponds to the generating function $\mathcal{R}(z)$. }
\label{fig:Rz0}
\end{figure}
While the solution to Eq.~(\ref{eq:P}) in the closed form may be hard to obtain, the leading singularity of the generating function $\mathcal{P}(z)$ can be often identified, revealing the tail behavior of the cluster-size distribution $P(s)$~\cite{lee04}.  According to Eq.~(\ref{eq:R}), the inverse of the generating function $\mathcal{R}(z)$ is represented as
\begin{equation}
z(\mathcal{R})  = {\mathcal{R} \over h(\mathcal{R})},
\label{eq:zR}
\end{equation}
and some examples are shown in Fig.~\ref{fig:Rz0}. Once the singularity of $\mathcal{R}(z)$ is identified, one can obtain that of $\mathcal{P}(z)$ by using Eq.~(\ref{eq:P}). 

We are  interested in the asymptotic behavior of the cluster-size distribution near and at the critical point for $\lambda<\lambda_l$ and $\lambda>\lambda_l$, respectively. Let us first consider the cluster-size distribution for small $K$ and  $\lambda<\lambda_l$, for which the critical point is zero, i.e., $K_c=0$.  When $K$ is smaller than $ K_1(\lambda)$  in Eq.~(\ref{eq:K1}), $h(z)$ is given by Eq.~(\ref{eq:g1z1}). As shown in Fig.~\ref{fig:Rz0}, there is a point $(\mathcal{R}_0,z_0 = \mathcal{R}_0 /h(\mathcal{R}_0))$ where the inverse function $z(\mathcal{R})$ has zero derivative. Around the point,  it is expanded as 
\begin{align}
z = z_0 - {1\over 2} {\mathcal{R}_0 h^{\prime\prime}(\mathcal{R}_0) \over h(\mathcal{R}_0)^2} (\mathcal{R}-\mathcal{R}_0)^2 + \cdots,
\label{eq:zRk0expand}
\end{align}
where $h(x)$ is given in Eq.~(\ref{eq:g1z1}) and $h^{\prime\prime}(x) = (d^2/dx^2) h(x)$.  
Therefore $\mathcal{R}(z)$ possesses a square-root singularity around $z_0$ as 
\begin{equation}
\mathcal{R}(z) \simeq \mathcal{R}_0 - \sqrt{2 h(\mathcal{R}_0)^2 \over \mathcal{R}_0 h^{\prime\prime}(\mathcal{R}_0)} \sqrt{z_0-z} + \cdots.
\end{equation}
Expanding $\mathcal{P}(z)=z g (\mathcal{R}(z))$, as in Eq.~(\ref{eq:P}), around $z_0$, we find that 
\begin{align}
\mathcal{P}(z) \simeq z_0 g (\mathcal{R}_0) - 2 K \sqrt{2 \mathcal{R}_0 h(\mathcal{R}_0)^2 \over h^{\prime\prime}(\mathcal{R}_0)} \sqrt{z_0 - z} + \cdots.
\end{align}
Recalling that $\mathcal{P}(z) = \sum_s P(s) z^s$ and using the relation 
\begin{align}
(1-x)^\theta &= \sum_{s=0}^\infty {(-x)^s \over s!} {\Gamma(\theta+1) \over \Gamma(\theta -s+1)} \nonumber\\
&=  -\sum_{s=0}^\infty {x^s \over s!} {\theta! (s-\theta-1)! \sin \pi\theta \over \pi},
\label{eq:tauberian}
\end{align}
which allows us to use  the Stirling's formula $s! \simeq s^s e^{-s} \sqrt{2\pi s}$ for large $s$, we obtain the tail behavior of $P(s)$ as
\begin{align}
P(s) &\simeq 
2K \, p_{\rm (I)} s^{-{3\over 2}} e^{-{s\over s_0}},
\label{eq:Ps1}
\end{align}
where  
$p_{\rm (I)} =  \sqrt{ z_0 \mathcal{R}_0 h(\mathcal{R}_0)^2 \over 2 \pi h^{\prime\prime}(\mathcal{R}_0)}$
and $s_0 = {1\over \ln z_0} = {1\over \ln \left({\mathcal{R}_0 \over h(\mathcal{R}_0)}\right)}$ are constants depending on $\lambda$.  

\begin{figure}
\includegraphics[width=\columnwidth]{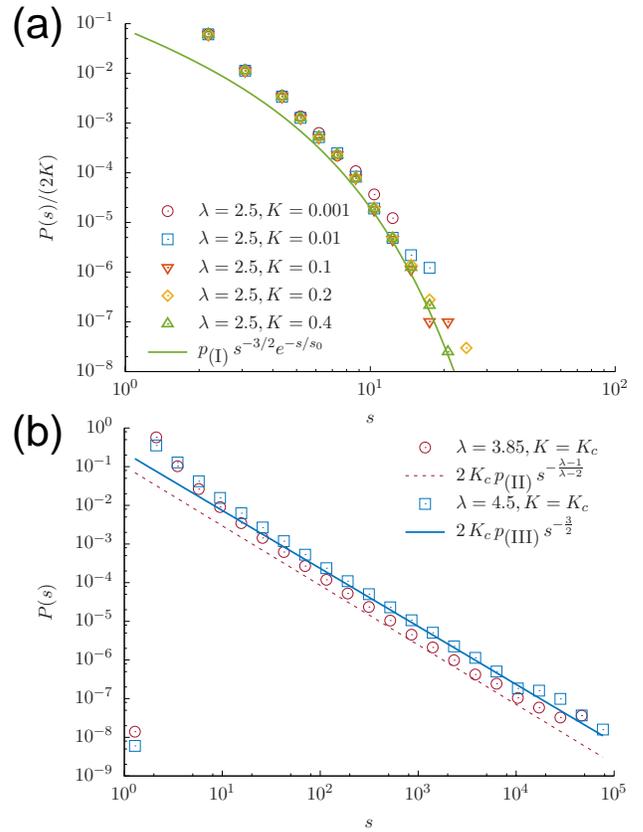}
\caption{ Cluster-size distribution $P(s)$ near or at the critical point in the configuration-model PL graphs of $N=10^7$ nodes and several degree exponents. (a) Plots of $P(s)/(2K)$ versus $s$  for $\lambda=2.5$ and $K=0.0001, 0.01, 0.1, 0.2$, and $0.4$. The theoretical prediction from Eq.~(\ref{eq:Ps1}) is also shown (line). (b) Plots of $P(s)$ versus $s$ at the critical point $K_c =0.628$ and $K_c=0.709$    for $\lambda=3.85$ and $\lambda=4.5$, respectively.  The lines are the theoretical predictions from Eq.~(\ref{eq:Ps23}). }
\label{fig:ps}
\end{figure}

Note that the cut-off constant $s_0$ is finite for $\lambda<\lambda_l$ and diverges at $\lambda = \lambda_l$. 
The exponential decay of $P(s)$  for any nonzero $K$ and $\lambda<\lambda_l$ implies that our configuration-model graph is in the supercritical (percolating) phase.  Moreover, $P(s)$ is independent of $K$ as long as $0<K<K_1(\lambda)$ for given $\lambda$. This is highly contrasted to $P(s)$ in the static model, which decays as a power-law for a wide range of $s$ depending on $K$, as in Eq.~(\ref{eq:Ps3_static}), if $K$ is small and $2<\lambda<3$~\cite{lee04}. The invariance of $P(s)/(2K)$ in Eq.~(\ref{eq:Ps1})  against the variation of $K$ in the range $0<K<K_1(\lambda)$, as confirmed numerically in Fig.~\ref{fig:ps} (a), originates in the specific form of the degree distribution for $0<\xmin<1$:
\begin{equation}
\pdegconfig_\xmin (k) = 
\begin{cases}
1 - {\langle k\rangle \over \zeta(\lambda-1)} & {\rm for} \ k=0\\
\langle k\rangle {k^{1-\lambda} - (k+1)^{1-\lambda} \over \zeta(\lambda-1)}& {\rm for} \ k\geq 1
\end{cases}
,
\label{eq:Pkx0}
\end{equation}
which leads $(k+1) \pdegconfig_\xmin (k+1)/\langle k\rangle$ for $k\geq 0$ and its generating function $h(z)$ to be independent of $K$ or $\xmin$ underlying the plateaus of $h'(1)$ in Fig.~\ref{fig:g11}. 

At  the critical point $K_c$ for $\lambda\geq\lambda_l$,  the inverse function $z(\mathcal{R})$ in Eq.~(\ref{eq:zR}) should be computed with $h(x)$ given in Eq.~(\ref{eq:g1z2}) since $\kmin=1$ at the critical point. Then one finds $z(\mathcal{R})$ has zero derivative at $\mathcal{R}_0=1$.  Using Eq.~(\ref{eq:g1zexpand}) at $K=K_c$, where $c_1=-1$, we find that $z(\mathcal{R})$ is expanded around $(\mathcal{R}_0,z_0)=(1,1)$ as 
\begin{align}
z = 1 - {c_2-1\over 2} (1-\mathcal{R})^2  - {c_{\lambda-2} \over (\lambda-2)!} (1-\mathcal{R})^{\lambda-2} + \cdots.
\label{eq:zRk1expand}
\end{align}
In the right-hand side of Eq.~(\ref{eq:zRk1expand}), the $(1-\mathcal{R})^2$ term is dominant over $(1-\mathcal{R})^{\lambda-2}$ for $\lambda>4$ and the latter is dominant for $\lambda_l\leq \lambda<4$. 
By the relation between $\mathcal{P}(z)$ and $\mathcal{R}(z)$ in Eq.~(\ref{eq:P}) and the expansion $g(z) = 1 - 2K (1-z) + O((1-z)^2, (1-z)^{\lambda-1})$, we find that $\mathcal{P}(z)$ behaves around $z=1$ as 
\begin{align}
\mathcal{P}(z) &\simeq  
\begin{cases}
1 - 2K \left((\lambda-2)! \over c_{\lambda-2}\right)^{1\over \lambda-2} (1-z)^{1\over \lambda-2} & {\rm for} \ \lambda_l\leq \lambda<4,\\
1 - 2K \sqrt{2\over c_2-1} \sqrt{1-z} + \cdots &{\rm for} \ \lambda>4.
\end{cases}
\end{align}
Finally we obtain the tail behavior of the cluster-size distribution using Eq.~(\ref{eq:tauberian}), which is characterized by the exponent ${\lambda-1\over \lambda-2}$ and $3/2$ for $\lambda_l\leq \lambda<4$ and $\lambda>4$, respectively, as 
\begin{align}
P(s)\simeq 
\begin{cases}
2K_c \, p_{\rm (II)}  s^{-{\lambda-1 \over \lambda-2}} & {\rm for} \ \lambda_l\leq \lambda<4,\\
2K_c \, p_{\rm (III)} s^{-{3\over 2}} & {\rm for} \ \lambda>4,
\end{cases}
\label{eq:Ps23}
\end{align}
with the coefficients given by
$p_{\rm (II)} = \left((\lambda-2)! \over c_{\lambda-2}\right)^{1\over \lambda-2} { \sin \left({\pi \over \lambda-2}\right) \over \pi} \left({1\over \lambda-2}\right)!$ and
$p_{\rm (III)} = \sqrt{1\over 2 \pi (c_2 -1)}$.
While the exponent depends on $\lambda$ for $3<\lambda<4$ in the static model, it does only for $\lambda_l\leq \lambda<4$ in the configuration model.

In Fig.~\ref{fig:ps}, we present the theoretical results in Eqs.~(\ref{eq:Ps1}) and (\ref{eq:Ps23}), including the coefficients $p_{\rm (I)}, p_{\rm (II)}$, and $p_{\rm (III)}$, along with the simulation results for the cluster-size distributions for selected values of the degree exponent, which are in good agreement  regarding their tail behaviors.


\section{Giant component in the supercritical regime: Importance of low-degree nodes}
\label{sec:low}

\begin{figure}
\includegraphics[width=\columnwidth]{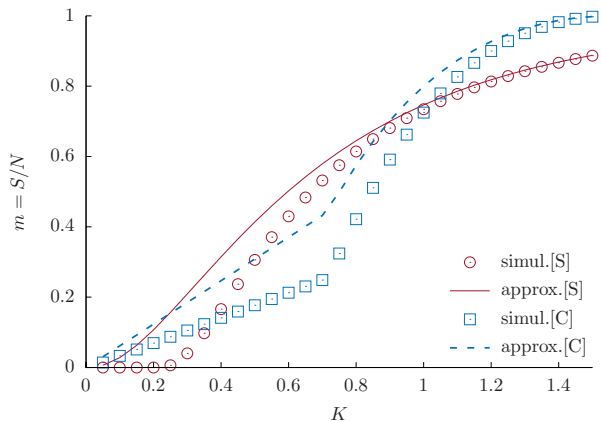}
\caption{Plots of the relative size of the giant component $m$ versus the number of links per node $K$ in the static model ([S]) and the configuration model ([C]) with $N=10^6$ and $\lambda=3.3$. Points are from the exact solution to Eq.~(\ref{eq:mu}) and lines are from the approximation in Eq.~(\ref{eq:m_approx}). }
\label{fig:SNcompare}
\end{figure}

The difference of the giant component size $m$ as a function of the number of links per node $K$ between our configuration model and the static model  is the most dramatic when the degree exponent $\lambda$ is between $3$ and $\lambda_l\simeq 3.81$; The critical point $K_c$ is nonzero for the static model while it is zero for the configuration model. 
Actually  $m$ is widely different in the whole range of $K$ between the two models.
 In  Fig.~\ref{fig:SNcompare}, it is shown  that $m$ is  larger in the configuration-model graph if $K$ is either very small or large, while it is larger in the static model in the intermediate range of $K$.  

The excellent agreement between simulations  and analytic results, derived based on the degree distribution only, means that  such different behaviors of the percolation transition 
and the  giant component size in the whole range of $k$ between the two models stem from different forms of their degree distributions.  
Examining the functional forms of  $\pdegstatic(k)$ and $\pdegconfig_\xmin(k)$  given in Eqs.~(\ref{eq:Dkstatic}) and (\ref{eq:Dkconfig}) and examples in Fig.~\ref{fig:KxminPk},  one 
finds that they share the same asymptotic behaviors for large $k$ characterized by the same degree exponent but behave differently in the small-$k$ region. This suggests the relevance of  the low-degree behavior of the degree distribution to the size of the giant component in graphs. 

The importance of low-degree nodes is understood also in computing the giant component size for large $K$.  It has been shown ~\cite{Deijfen2018} that the bounds of the giant component size in the supercritical regime are essentially determined by the low-degree behavior of the degree distribution. 
Here we present an approximate expression for the size of the giant component when $K$ is very large, which helps us better understand the different giant component sizes between the two models in the supercritical regime. Assuming that $K$ is large, we find $u$ from Eq.~(\ref{eq:mu}) expanded  for the degree distribution $D(k)$ as 
\begin{equation}
u=  {\pdegconfig (1) \over 2K} + {\pdegconfig(1) \pdegconfig(2) \over 2K^2} + O(K^{-3}),
\end{equation}
and the giant component has relative size $m$ given by 
\begin{align}
m_{\rm approx} &\simeq 1 - \pdegconfig(0) - \pdegconfig(1) u - \pdegconfig(2) u^2  \nonumber\\
&\simeq 1 - \pdegconfig(0) - {\pdegconfig(1)^2 \over 2K} - {3\pdegconfig(1)^2 \pdegconfig(2)\over 4K^2}.
\label{eq:m_approx}
\end{align}
While obtained for $K$ large, this approximation works very well for $K\gtrsim 1$, and reasonably well even for $K$ small, in both the static and the configuration model (Fig.~\ref{fig:SNcompare}). Moreover, as noted above, the static model will form a larger giant component than the configuration model for intermediate values of $K$, 
for example, $0.4\lesssim K\lesssim 1$ for $\lambda=3.3$ in Fig.~\ref{fig:SNcompare},  and we observe the same phenomenon for $m_{\rm approx}$ in the range $0.3\lesssim K\lesssim 0.9$. This suggests that the difference of $\pdegconfig(k)$ for small $k$ such as $k=0,1$, or $2$ between the two models is partly responsible for their difference in $m$. 
As shown in Appendix~\ref{sec:real}, the approximation in Eq.~(\ref{eq:m_approx}) is also useful in estimating  the giant component size of real-world networks while they lose links. Therefore our results demonstrate that both the large- and small-$k$ behavior of the degree distribution is important for understanding and controlling the global connectivity of complex networks.


\section{Discussion}
\label{sec:summary}

We have shown that the full functional form of the degree distribution can control the percolation transition and critical phenomena on PL graphs. The exponent characterizing the PL decay of the degree distribution, which has received most attention in theoretical and empirical settings, may not be sufficient to predict such behaviors. We have demonstrated this point on PL networks generated with the configuration model equipped with a generalized PL degree distribution where the number of links and the degree exponent can be tuned separately. By studying the percolation transition in these networks numerically and analytically, and comparing its outcomes to known results illustrated by the static model~\cite{lee04}, we have shown in detail how far different functional forms of the degree distributions sharing the same degree exponent may alter the critical phenomena. 

In previous studies, the role of diverging moments was shown to be important across models and dynamics,  from percolation to other phenomena on PL networks such as disease spreading and synchronization.  
Likewise,  we propose that  nodes with low degree may also wield a general influence on critical behaviors, which should be explored. A better understanding of whether and when the lower range of the degree distribution controls critical and general dynamical properties  would prove beneficial for a wide range of studies and applications. 

Our proposed degree distribution exhibits, for parameter values $0<\xmin<1$ in Eq.~(\ref{eq:Pkx0}), a PL shape across the largest range of degrees $k$.  It possesses a special property of invariance: the probability of being connected to a node with $k$ links (computed as $k D(k)/\langle k\rangle$) does not depend on the average degree $\langle k \rangle$. Therefore, critical behaviors become independent of the number of links in the network, and we expect this property to translate to similarly robust phenomena in other dynamical processes.

\begin{acknowledgments}
This work was supported by 
National Research Foundation of Korea (NRF) grants funded by the Korean government (Grant No. 2019R1A2C1003486) and
an Inha University Research Grant (No. 59212).
\end{acknowledgments}



\appendix

\section{Percolation transition in the static model}
\label{sec:static}
 
 The static model is a generalization of the Erd\"{o}s-R\'{e}nyi graph~\cite{Erdos:1960} to an asymptotic PL degree distribution. Different pairs of nodes are connected with different probabilities but independently. Therefore the graphs obtained by removing links randomly and independently in a PL graph are similar to the static-model graphs, which is further discussed in Appendix~\ref{sec:factorial}.  This similarity holds notably for  the degree distribution and the critical phenomena associated with the percolation transition.   Moreover, due to the independence of  connecting different  pairs of nodes,  one can obtain analytically the giant component and the size distribution of the finite clusters with the help of the Potts model formulation~\cite{lee04}.    Here we summarize the important properties of the static-model  graphs, as their comparison with the results for our configuration model is  of main concern in this paper. 

The relative size $m$ of the LCC  in the static model exhibits a transition, from zero to a nonzero value as a function of $K$ at a threshold $K_c$  if the degree exponent $\lambda$ is larger than the lower-critical degree exponent $\lambda_l=3$~\cite{PhysRevLett.85.4626,lee04}. When $2<\lambda<\lambda_l=3$, $K_c=0$ and thus no transition occurs, and $m$  grows superlinearly with $K$ in the small-$K$ regime. Such different critical behaviors for  $\lambda$ below and above $\lambda_l=3$ have been recognized as  the most remarkable feature of critical phenomena on PL graphs,  originating in the diverging second moment of  the PL degree distribution for $\lambda\leq 3$.  

The percolation threshold $K_c$ in the static model is given by~\cite{lee04} 
\begin{equation}
K_c = 
\begin{cases}
0 & {\rm for } \ 2<\lambda\leq 3 \\
{(\lambda-1)(\lambda-3) \over 2(\lambda-2)^2} & {\rm for} \ \lambda>3
\end{cases}
.
\label{eq:Kc_static}
\end{equation}
The relative size $m$ of the LCC is zero for $K<K_c$ and 
\begin{align}
m  \sim \left({K\over K_c} -1 \right)^\beta 
\label{eq:m_static}
\end{align}
for $K\geq K_c$ if the degree exponent is as large as $\lambda>3$. Here the critical exponent $\beta$ is given by 
\begin{equation}
\beta  = 
\begin{cases}
{1\over \lambda-3} & {\rm for} \ 3<\lambda<4\\
1 & {\rm for} \ \lambda>4
\end{cases}
.
\label{eq:beta12_static}
\end{equation}
For $2<\lambda<3$, the LCC size behaves as 
\begin{align}
m  \sim K^{1\over 3-\lambda}  \ {\rm for} \ K\ll 1.
\label{eq:beta3_static}
\end{align}

At $K=K_c$ for $\lambda>3$, the cluster-size distribution $P(s)$ takes a PL form as 
\begin{equation}
P(s) \sim s^{1-\tau}
\label{eq:Ps12_static}
\end{equation}
with the critical exponent 
\begin{equation}
\tau  = 
\begin{cases}
{2\lambda-3\over \lambda-2} & {\rm for} \ 3<\lambda<4\\
{5\over 2} & {\rm for} \ \lambda>4
\end{cases}
.
\label{eq:tau12_static}
\end{equation}
For $2<\lambda<3$, $P(s)$ for small $K$ (near the zero critical point) behaves as 
\begin{equation}
P(s)\sim \left({s\over K}\right)^{1-\lambda}.
\label{eq:Ps3_static}
\end{equation}

In the removal of randomly selected links reducing $K$  in a  PL graph, the degree distribution maintains its asymptotic PL behavior, and thus the shrinkage and extinction  of the giant component is expected to be characterized by the above results,   as different pairs of nodes are treated independently in the static model. 
The degree distribution of the static-model PL  graphs  with $f$ fraction of links removed randomly is equal to that of the static model with $K'=K(1-f)$, as  shown in Appendix~\ref{sec:factorial}.

The absence of a critical threshold for $2<\lambda<3$ and the critical exponents continuously varying with the degree exponent $\lambda$ are observed in a wide range of dynamical processes including epidemic spreading~\cite{PhysRevLett.86.3200}, Ising model~\cite{PhysRevE.66.016104}, synchronization~\cite{PhysRevE.70.026116,PhysRevE.72.026208}, order-disorder transition in the boolean dynamics~\cite{LEE2007618,lee08jpa}, etc. For instance, the critical exponent $\beta$ for the Ising model and the synchronization order parameter in the Kuramoto model on PL graphs is also given by $\beta = 1/2$ for $\gamma>5$ and $1/(\lambda-3)$ for $3<\lambda<5$.

\section{The largest degree in the configuration-model PL networks}
\label{sec:uncorr}

In the configuration-model PL networks, the density of self or multiple connections is negligible if $\lambda>3$ or the maximum degree cutoff $k_{\rm max}\sim \sqrt{N}$ is introduced for $2<\lambda<3$~\cite{PhysRevE.71.027103}. In simulations, we restrict the range of $x$ to $x\in [\xmin, x_{\rm max}]$ in Eq.~(\ref{eq:pareto_xmin}) so as to realize the upper cutoff $k_{\rm max} \sim \sqrt{N}$ for $2<\lambda<3$. However, the introduction of $k_{\rm max}$ does not significantly change any of the presented theoretical results in the limit $N\to\infty$, and so we will use Eq.~(\ref{eq:pareto_xmin}) for simplicity in the theoretical analysis.

\section{Scaling of the factorial moments in the static model and under random removal of links}
\label{sec:factorial}

The $r$-th factorial moment of a degree distribution $D(k)$ is defined as 
\begin{equation}
\langle (k)_r\rangle \equiv \langle k (k-1) (k-2) \cdots (k-r+1) \rangle 
\label{eq:factorial_def}
\end{equation}
and can be evaluated by differentiating the generating function $g(z) = \sum_{k=0}^\infty D(k) z^k$ $r$ times at $z=1$ as 
\begin{equation}
\langle (k)_r\rangle = {d^r \over dz^r} g(z)\bigg|_{z=1}.
\label{eq:factorial_diff}
\end{equation}
Here we show that when links are added randomly  (the static model) or removed randomly,  the factorial moments $\langle (k)_r\rangle$ scale with the mean degree $\langle k\rangle=\sum_k k D(k)$ as $\langle (k)_r\rangle \sim \langle k\rangle^r$. 

In the static model~\cite{goh01}, the generating function of the degree distribution $g(z) \equiv \sum_k \pdegconfig(k) z^k$ of the static model graphs of $N$ nodes, $L$ links, and degree exponent $\lambda$ is given by~\cite{lee04}
\begin{align}
g(z) &=  {1\over N} \sum_{i=1}^N \prod_{j\ne i} \left[ e^{-2L w_i w_j} + z  \left(1-e^{-2L w_i w_j} \right)\right] \nonumber\\
&=\tilde{\Gamma} (2 K (1-z)),
\label{eq:g0z_static}
\end{align}
where $w_i = {i^{-{1\over\lambda-1}}\over \sum_{\ell=1}^N \ell^{-{1\over \lambda-1}}}$ is the probability of node $i$ to be selected to gain a link and the function $\tilde{\Gamma}(y)$ is given by 
\begin{equation}
\tilde{\Gamma}(y) = (\lambda-1) \left({\lambda-2\over \lambda-1} y \right)^{\lambda-1} \Gamma\left(1-\lambda, {\lambda-2 \over \lambda-1} y\right)
\label{eq:Gammatilde}
\end{equation}
with $\Gamma(s,x)$ the incomplete Gamma function $\Gamma(s,x)\equiv \int_x^\infty dt \, t^{s-1} e^{-t}$.  As $g(z)$ is a function of $2K(1-z)$, the $r$-th derivative of $g(z)$ is proportional to $K^r$ and we have 
\begin{equation}
\langle (k)_r\rangle = (2K)^r \tilde{\Gamma}^{(r)} (0)
\label{eq:kr_static}
\end{equation}
for $r<\lambda-1$. Here $\tilde{\Gamma}^{(r)}(0) = {d^r \over dy^r} \tilde{\Gamma}(y)\bigg|_{y=0}$. When $r>\lambda-1$, $\tilde{\Gamma}^{(r)}(0)$ diverges.  In Figs.~\ref{fig:KxminPk}(c) and~\ref{fig:KxminPk}(d),  it is shown for the static model that $\langle (k)_2\rangle$ is proportional to $K^2$ and $\langle (k)_3\rangle$ is to $K^3$. 

A similar scaling relation to Eq.~(\ref{eq:kr_static}) holds also for graphs with links removed randomly. Consider a graph of $K_0$ links per node, a degree distribution $D_0(k)$. When a fraction  $f$ of links are removed randomly, the number of links per node is given by 
\begin{equation}
K = K_0 (1-f),
\label{eq:Kf}
\end{equation} 
and the degree distribution  is changed to 
\begin{equation}
\pdegconfig_f(k) = \sum_{k'=k}^\infty \pdegconfig(k') \binom{k'}{k} f^{k'-k} (1-f)^k.
\end{equation}
The generating function  is then given by 
\begin{equation}
g_f(z) = g(f+(1-f)z)  = \tilde{g}(2K(1-z))
\label{eq:g0z_link}
\end{equation}
with $\tilde{g}(x) = g(1- {x\over 2 K_0})$. Therefore the factorial moments scales with respect to $K$ in the same way as in Eq.~(\ref{eq:kr_static}):
\begin{equation}
\langle (k)_r \rangle = (2K)^r \tilde{g}^{(r)}(0)
\label{eq:kr_link}
\end{equation}
with $\tilde{g}^{(r)}(0) = {d^r \over dy^r} \tilde{g}(y) \bigg|_{y=0}$.

 \section{Non-random link-removal process preserving the form of the new degree distribution}
\label{sec:linkremoval}

\begin{figure}
\includegraphics[width=0.9\columnwidth]{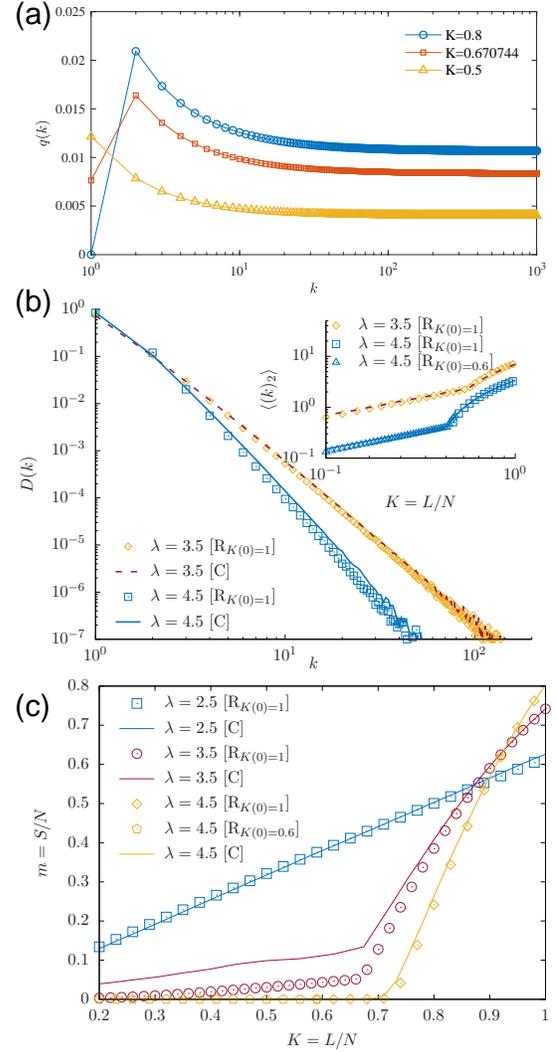}
\caption{
Properties of the graphs obtained by the nonrandom link-removal process. (a) Plots of $q(k)$ versus $k$ for $\lambda=3.5$, $q_o=0.01$, and different $K$. (b) The degree distributions of the link-removed graphs ([R]) with $K=0.6$.  Links are removed in the initial configuration-model graphs of $K(0)=1$. The degree distributions of the configuration-model graphs ([C]) for the same value of $K (=0.6)$ are also shown. Inset: The second factorial moments $\langle (k)_2 \rangle$ of the link-removed graphs  (points) and  our configuration-model graphs (lines).  For $\lambda=4.5$, the link-removed graphs from the configuration-model  graphs with $K(0)=1$ and $K(0)=0.6$ are used to cover the whole considered range of $K$. (c) The relative size of the LCC versus $K$ for the link-removed graphs and the configuration-model graphs.
}
\label{fig:linkremoval}
\end{figure}

The decrease of $K$ in our configuration-model graphs can be realized by a nonrandom link-removal process. Let us construct a configuration-model PL graph of $N$ nodes, $L(0)$ links, and degree exponent $\lambda$ as in Sec.~\ref{sec:model_config} at time $t=0$. As time $t$ increases,  it loses links, resulting in the decrease of $L(t)$, $K(t)$, and $\xmin(t)$ related by Eq.~(\ref{eq:Kxmin}), while the degree distribution $D(k,t)$ is equal to $D_{\xmin(t)}(k)$ in Eq.~(\ref{eq:Dkconfig}). To be specific, the following steps  are taken:
\begin{enumerate}[(i)]
%
\item At time $t$, a link, say, $e_{ij}$ connecting nodes $i$ and $j$, is selected randomly.
\item The selected link is removed with probability dependent on the degrees of the end nodes 
\begin{equation}
\ell_{ij} = q(k_i) \, q(k_j),
\end{equation}
or remains with probability $1-\ell_{ij}$. 
\item Time is increased by $dt = L(t)^{-1}$.
\item (i)- (iii) are repeated. 
\end{enumerate}
Here $q(k)$ is given by 
\begin{align}
q(k) 
&=
\begin{cases}
0 & (k<\kmin)\\
q_o {1-\eta\over 1 - {n-1 \over 2K}}  {k^{-\lambda} \over k^{1-\lambda} - (k+1)^{1-\lambda}} & (k=\kmin) \\
q_o \left\{ { \eta  \over 1 - {\kmin \over 2K}}  + {1-\eta\over 1 - {\kmin-1 \over 2K}}   \right\}{k^{-\lambda} \over k^{1-\lambda} - (k+1)^{1-\lambda}}  & (k>\kmin)
\end{cases}
,
\label{eq:qk}
\end{align}
where $\kmin=\lfloor \xmin(t)\rfloor$ is the minimum degree at time $t$, $q_o$ is a constant controlling the rate of link removal, and $\eta$ is 
\begin{equation}
\eta = \min\{1,L(t) - N K_n\}
\label{eq:eta}
\end{equation}
with $K_n$ 
\begin{equation}
K_n ={1\over 2} \{ n + n^{\lambda-1} \zeta(\lambda-1,n+1)\}.
\end{equation}
corresponding to $\xmin=n$ for integer $n$ by Eq.~(\ref{eq:Kxmin}), and generalizing Eq.~(\ref{eq:K1}). It should be noted that $\eta$ is $1$ and $q(\kmin)=0$ unless $L(t)-1< N K_n< L(t)$ for some integer $n$.  

The link-removal probability $\ell_{ij}$ is not uniform but depends on the degree of the end nodes via the function $q(k)$ in Eq.~(\ref{eq:qk}), the behaviors of which are shown in Fig.~\ref{fig:linkremoval} (a). In the most period of time,  $\eta=1$ and $q(\kmin)=0$, implying that the links incident on the nodes of the minimum degree cannot be removed. They can be removed only when the minimum degree  $\kmin(t) = \lfloor \xmin\rfloor$ will be changed by the removal of a single link, for which there exists an integer $n$ such that $L(t)-1< N K_n< L(t)$ and thus $\eta<1$. As shown in Fig.~\ref{fig:linkremoval} (b), the graphs obtained by this link-removal process have the same degree distribution, in the form of  Eq.~(\ref{eq:Dkconfig}), and the same moments as our configuration-model graphs for given $K$. 

Below we explain how to derive $q(k)$ in Eq.~(\ref{eq:qk}).  By the link-removal process presented above, the number of links per node decreases with time as 
\begin{equation}
{d K\over dt}  = -{1\over N} \sum_{i<j} A_{ij} \ell_{ij}= -K \langle q\rangle^2,
\label{eq:dKdt}
\end{equation} 
with 
\begin{align}
\langle q\rangle &= 
 \sum_{k} {k D(k,t) \over \langle k\rangle} q(k),
 \label{eq:qave}
\end{align}
where the approximation ${\sum_{i,j} A_{ij} f(k_j)g(k_j) \over \sum_{ij} A_{ij}} = \sum_{k,k'} {k D(k,t) k' D(k',t) \over \langle k\rangle^2} f(k)  g(k')$ is used,  assuming that  there is no degree-degree correlation between adjacent nodes. $\langle q\rangle$ will be shown later to be equal to the constant $q_o$.   The degree distribution $D(k,t)$ evolves with time $t$ as 
\begin{equation}
{\partial \over \partial t} D(k,t) =  v(k+1) D(k+1,t) - v(k) D(k,t),
\label{eq:Ddt}
\end{equation}
where $v(k)$ is the fraction of the nodes losing one link among the nodes of degree $k$, evaluated by  
\begin{equation}
v(k) = k q(k) \langle q\rangle,
\label{eq:vq}
\end{equation}
and we assume that no node loses more than one link in the time interval $dt$, holding if $\max_{ij} k_i \ell_{ij} dt \ll 1$.  

Suppose that $D(k,t)$ is equal to $D_{\xmin(t)}(k)$ in Eq.~(\ref{eq:Dkconfig}). Then, from Eq.~(\ref{eq:Ddt}), one finds that $v(k)$ is related to the cumulative degree distribution $F_{\xmin(t)} (k)$ in Eq.~(\ref{eq:pareto_xmin}) as 
\begin{equation}
v(k)D_{\xmin(t)} (k)  = \sum_{k'=\kmin-1}^{k-1} {\partial \over \partial t} D_{\xmin(t)} (k')   =  - {\partial \over \partial t} F_{\xmin(t)} (k),
\label{eq:vDF}
\end{equation}
where we used $F_{\xmin}(k) = \sum_{k'=k}^\infty D_{\xmin}(k') =  1 - \sum_{k'=\kmin}^{k-1} D_{\xmin}(k')$,  and $v(\kmin-1)=0$. 
In implementing numerically the link-removal process,  we deal with finite systems, for which the decrease of $K$ cannot be smaller than $1/N$ and thus a small but finite time interval $\Delta t$ should be considered. When the minimum degree will not be changed but fixed at $\kmin=n$ by the removal of a single link,  or equivalently $\eta = 1$ in Eq.~(\ref{eq:eta}),  the time dependence of $F_{\xmin(t)}(k)$ in Eq.~(\ref{eq:pareto_xmin}) arises solely from $\xmin(t)$,
\begin{align}
{\partial \over \partial t} F_{\xmin(t)}(k)\bigg|_{\kmin=n} &= {d (\xmin^{\lambda-1}) \over dt} {\partial \over \partial (\xmin^{\lambda-1})} F_{\xmin}(k)\bigg|_{\kmin=n}\nonumber\\
&=
\begin{cases} 
0 & (k\leq n)\\
 {2{dK\over dt} \over \zeta(\lambda-1,\kmin+1)  } k^{1-\lambda} & (k>n)
\end{cases}
,
\label{eq:Ddt1}
\end{align}
where we used $F_{\xmin}(\kmin)=1$ for $\xmin>\kmin$, and 
 Eq.~(\ref{eq:Kxmin}). Using Eq.~(\ref{eq:dKdt}), one obtains  Eq.~(\ref{eq:qk}) with $\eta=1$ and $q_o=\langle q\rangle$.

Suppose that the removal of a single link at time $t$ will result in changing  $\kmin$ from $n$ to $n-1$.  Such a decrease of $\kmin$ causes $D_{\xmin}(k)$ to have newly a nonzero value at $k=n-1$, which should be taken care of in the numerical implementation of the derivative of $F_{\xmin}(k)$ in Eq.~(\ref{eq:vDF}).  Assuming that $L(t)$ decreases linearly with time in the time interval between  $t$ and $t+\Delta t$, we find that $\kmin=n$ first for $\eta \Delta t$ and then $\kmin=n-1$ for $(1-\eta) \Delta t$ with $\eta$ in Eq.~(\ref{eq:eta}). Therefore $v(k)$ in this time interval should be evaluated as $v(k) = \eta  v_{n}(k)+ (1-\eta) v_{n-1}(k)$, where 
\begin{align}
v_{n}(k)  &= {-{\partial \over \partial t} F_{\xmin(t)} (k)|_{\kmin=n} \over  D_{\xmin(t)}(k)|_{\kmin=n}}, \nonumber\\
v_{n-1}(k) &= {-{\partial \over \partial t} F_{\xmin(t)} (k)|_{\kmin=n-1} \over D_{\xmin(t)}(k)|_{\kmin=n-1}},
\end{align}
with ${\partial \over \partial t} F_{\xmin(t)} (k)$ given in Eq.~(\ref{eq:vDF}). This leads to Eq.~(\ref{eq:qk}) for $0<\eta<1$ with $q_o = \langle q\rangle= \eta \langle q\rangle_{\kmin=n} + (1-\eta) \langle q\rangle_{\kmin=n-1}$.

In Fig.~\ref{fig:linkremoval} (c), we present the relative sizes of the LCC of the link-removed graphs and compare them with those from the configuration-model graphs. Their behaviors as functions of $K$ are  in good agreement for $\lambda=2.5$ and $4.5$. Qualitative agreement is also observed for $\lambda=3.5$, but  $m$ remains smaller in the link-removed graphs than in the configuration model. The origin of this deviation is not clear to us.  It is perhaps related to the degree-degree correlation that we neglected in the branching process approach but is generated during the link-removal process.  Also  the portion of removable links is found to be smaller than expected, which results in leaving no removable links at $K\simeq 0.59$ for $\lambda=4.5$ when started from $K(0)=1$. It can be understood in terms of the degree-degree correlation as follows. Starting from $K(0)=1$ for $\lambda=4.5$, a link having an end node of degree  one cannot be removed as long as $K(t)>K_1 =  0.563367\ldots$ and thus $q(1)=0$. Then the portion of removable links is evaluated by  $1- {N D(1) \over L} + D(1,1)$, where $D(1,1)$ is the portion of the links connecting two degree-one nodes. While the portion of degree-one nodes is identical between the link-removed graphs and the configuration-model graphs,   $D(1,1)$ is smaller in the former than in the latter and the number of removable links is found to decrease rapidly with time and becomes zero at $K\simeq 0.59$ for $\lambda=4.5$~\cite{linkremoval_degcorr}.  It is why we use another initial configuration-model graph of $K(0)=0.6$ to remove links in and  thereby cover the whole considered range of $K$ for $\lambda=4.5$ in Figs.~\ref{fig:linkremoval}(b) and~\ref{fig:linkremoval}(c). 

\section{Behaviors of $Q(\lambda,\kmin)$ and $B(\lambda,n)$}
\label{sec:QB}

\begin{figure}
\includegraphics[width=0.8\columnwidth]{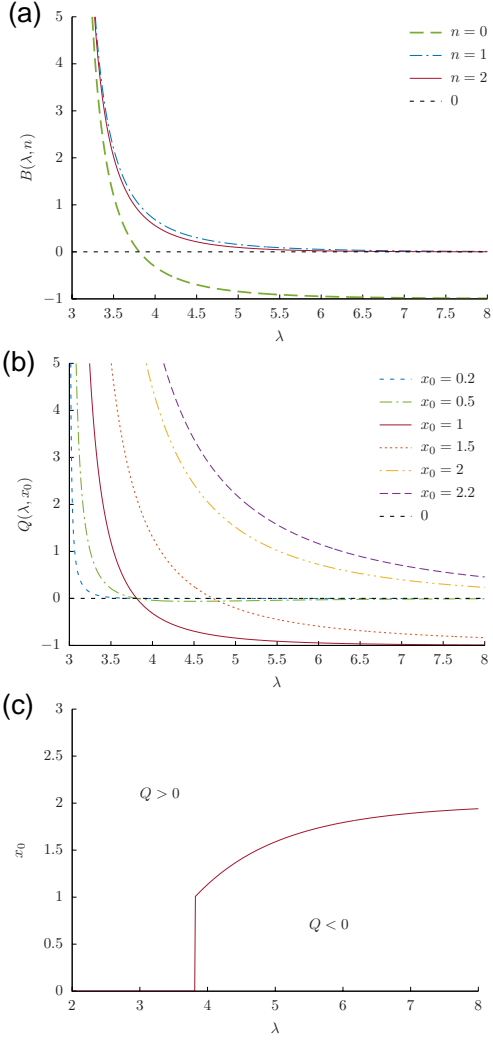}
\caption{Conditions for $Q>0$. (a) The behavior of $B(\lambda,n)$ as a function of $\lambda$ for different $n$'s. (b) $Q(\lambda,x_o)$ versus $\lambda$ for different $\xmin$'s. (c) The boundary between $Q>0$ and $Q<0$ in the $(\lambda,\xmin)$ plane drawn by Eq.~(\ref{eq:xminc}).}
\label{fig:xminc}
\end{figure}

Here we investigate the functional behaviors of $B(\lambda,n)$ and $Q(\lambda,\xmin)$ defined in Eq.~(\ref{eq:QB}), which are used to obtain the phase diagram. For very large $\lambda$, the function $B(\lambda,n)$ in Eq.~(\ref{eq:QB}) can be approximated as $B(\lambda,n)\simeq 2 (n+1)^{2-\lambda} - 3 (n+1)^{1-\lambda}  = (2n-1) (n+1)^{1-\lambda}$, which converges to 
\begin{equation}
\lim_{\lambda\to\infty} B(\lambda,n) = 
\begin{cases}
-1 & {\rm for} \ n=0 \\
0 & {\rm for} \ n\geq 1
\end{cases}
.
\label{eq:Binfty}
\end{equation}
For given $n$, $B(\lambda,n)$ decreases monotonically with $\lambda$, as its derivative is negative for all $\lambda$ and $n\geq 0$:
\begin{equation}
{\partial B(\lambda,n)\over \partial \lambda} = - \sum_{k=n+1}^\infty {2k-3 \over k^{\lambda-1}} \ln k <0.
\label{eq:Bdecrease}
\end{equation}
Note that $B(\lambda,n)$ diverges for $\lambda\leq 3$. For $\lambda>3$, we can refer to  Eqs.~(\ref{eq:Binfty}) and (\ref{eq:Bdecrease}) to find that $B(\lambda,n)$ is positive if $n$ is positive. For $n=0$, $B(\lambda,n)$ becomes negative for $\lambda>\lambda_l$ with $\lambda_l$ in Eq.~(\ref{eq:lambdac}). 
These behaviors of $B(\lambda,n)$ are shown in Fig.~\ref{fig:xminc} (a), which leads  $Q(\lambda,\xmin)$ to behave as in Fig.~\ref{fig:xminc} (b) by  Eq.~(\ref{eq:QB}). One finds that the value of $\lambda$ at which $Q(\lambda,\xmin)=0$ is fixed at $\lambda_l$ if $0<\xmin\leq 1$ and increases from $\lambda_l$ to infinity as $\xmin$ increases from $1$ to $2$. This boundary between $Q>0$ and $Q<0$ can be best represented by the critical line  $\xmin_c(\lambda)$ as a function of $\lambda$ for $\lambda\geq \lambda_l$  given in Eq.~(\ref{eq:xminc}) and another line $0<\xmin\leq 1$ at $\lambda=\lambda_l$, which 
are shown  in Fig.~\ref{fig:xminc} (c) and give the phase diagram in the plane ($\lambda,K$) in Fig.~\ref{fig:PD}.

\section{
Application of Eq.~(\ref{eq:m_approx}) for the giant component of  real-world networks under random link removal
}
\label{sec:real}

\begin{figure}
\includegraphics[width=0.75\columnwidth]{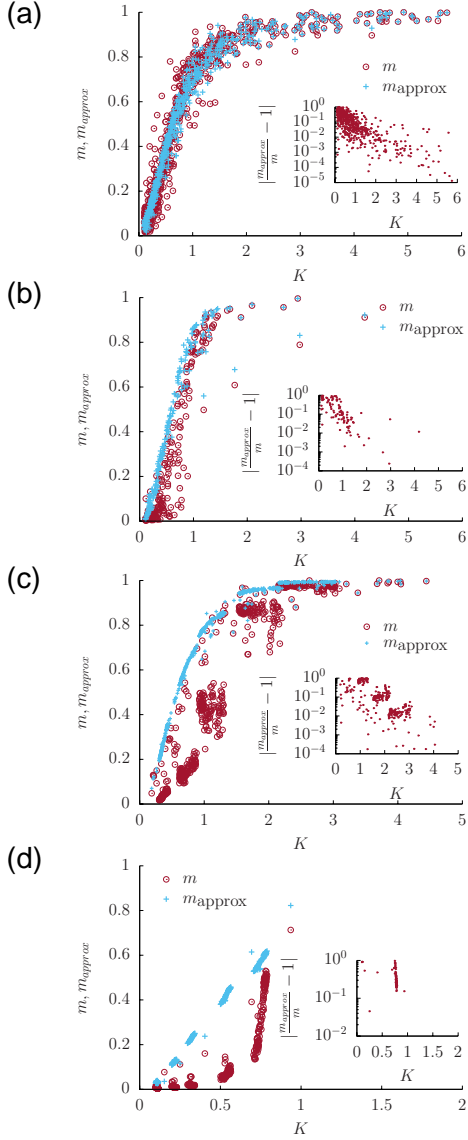}
\caption{
The relative size of the giant component $m$ and the approximation $m_{\rm approx}$   from  Eq.~(\ref{eq:m_approx})  versus the number of links per node $K$
in damaged networks, which are obtained by removing $q=0.3,0.4,0.5, 0.7,0.8$, and $0.9$ fraction of links randomly in  each of (a) 125 biological networks, (b) 40 technological, informational, and transportation networks, (c) 124 social networks, and (d) 112 economic networks. 
For given $q$ and each real-world network, 20 such damaged networks are generated and $m$ and $m_{\rm approx}$ are averaged over them.  In the insets, the relative difference of the two $|{m_{\rm approx} \over m} -1|$ is plotted versus $K$.}
\label{fig:empirical}
\end{figure}

The approximation in Eq.~(\ref{eq:m_approx}) allows us to estimate the size of the giant component in terms of the density of low-degree nodes, which can be of practical use. Suppose that a real network is being attacked,  losing a significant fraction of links. In such an emergency, it is important to know the size of the giant component. But the full adjacency matrix, necessary to identify the giant component and its size may be unavailable due to insufficient time or resources. Rather than struggling to collect information of the full adjacency matrix, one can instead count just the number of significantly damaged nodes such as those having zero, one or just two connected neighbors  and use them in  Eq.~(\ref{eq:m_approx}) to approximate the size of the giant component. 

To test this idea, we generate the damaged networks by removing randomly various fractions of links in real-world networks available in Ref.~\cite{empdata} and compute the relative sizes of the giant components as well as the densities of low-degree nodes. In Fig.~\ref{fig:empirical}, the approximation $m_{\rm approx}$ obtained by Eq.~(\ref{eq:m_approx}) shows a good agreement with $m$ for large $K$ in most of the real networks except for economic networks having relatively small $K$. The relative difference between $m$ and $m_{\rm approx}$ decreases with increasing $K$ as shown in the insets of Fig.~\ref{fig:empirical}. This suggests the usefulness of Eq.~(\ref{eq:m_approx}) in practical applications. Given the assumption of tree structure and negligible degree-degree correlation in the branching process approach leading to Eqs.~(\ref{eq:mu}) and (\ref{eq:m_approx}),  the agreement or deviation between $m$ and $\tilde{m}$ may be attributed to the validity or violation of the assumptions.

\begin{figure}
\includegraphics[width=0.75\columnwidth]{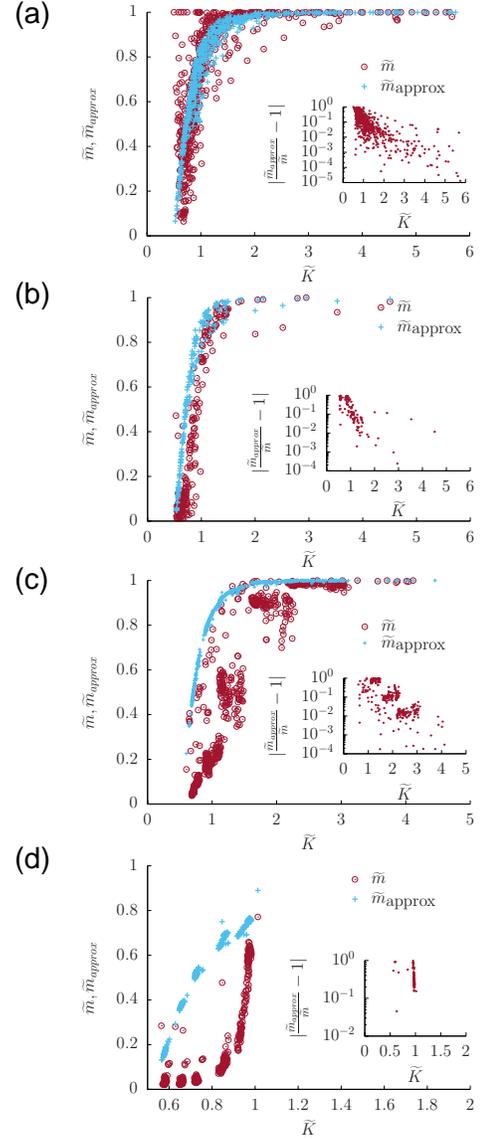}
\caption{
The relative size of the giant component $\tilde{m}$ and the approximation $\tilde{m}_{\rm approx}$   from  Eq.~(\ref{eq:mtilde_approx})  versus the rescaled number of links per node $\tilde{K}$ in the subgraph of nonisolated nodes.
The same damaged networks as considered in Fig.~\ref{fig:empirical} are used.
In the insets, the relative difference of the two $|{\tilde{m}_{\rm approx} \over \tilde{m}} -1|$ is plotted versus $\tilde{K}$.
}
\label{fig:empirical_tilde}
\end{figure}

Isolated nodes cannot belong to the giant component, and one might suspect that  the agreement between $m$ and $m_{\rm approx}$ in Fig.~\ref{fig:empirical} be driven by the first term in the right-hand-side of Eq.~(\ref{eq:m_approx}), $1-D(0)$, representing the portion of nonisolated nodes. If so, the relative size of the giant component in the subgraph of nonisolated nodes might be significantly different from the corresponding approximation from Eq.~(\ref{eq:m_approx}). To check this possibility, 
we consider  the relative size of the giant component in the {\it subgraph} of nonisolated nodes $\tilde{m} = {S\over \tilde{N}}$ as a function of its number of links per node $\tilde{K} = {L \over \tilde{N}}$, where   $\tilde{N}$ is the number of nonisolated nodes. As $\tilde{N}=N ( 1-D(0))$, one sees that $\tilde{m}= {m \over 1-D(0)}$ and $\tilde{K} = {K \over 1-D(0)}$. Similarly, the degree distribution of the subgraph is also given by $\tilde{D}(k) = {D(k)  \over 1-D(0)}$ for $k\geq 1$. 
The approximation for $\tilde{m}$ based on  Eq.~(\ref{eq:m_approx}) is therefore given by
\begin{equation}
{\tilde{m}}_{\rm approx} = {m_{\rm approx} \over 1-D(0)} = 1  - {\tilde{D}(1)^2 \over 2\tilde{K}} - {3 \tilde{D}(1)^2 \tilde{D}(2) \over 4 \tilde{K}^2}.
\label{eq:mtilde_approx}
\end{equation}
In Fig.~\ref{fig:empirical_tilde}, we compare $\tilde{m}$ and $\tilde{m}_{\rm approx}$ as functions of $\tilde{K}$, which show as good agreement as between $m$ and $m_{\rm approx}$ in Fig.~\ref{fig:empirical}.

%

\end{document}